\documentclass[12pt,preprint]{aastex}

\begin{document}

\title{Smooth and Starburst Tidal
Tails in the GEMS and GOODS Fields}

\author{Debra Meloy Elmegreen \affil{Vassar College, Dept. of Physics \& Astronomy, Box 745,
Poughkeepsie, NY 12604; elmegreen@vassar.edu}}
\author{Bruce G. Elmegreen\affil{IBM Research Division, T.J. Watson
Research Center, P.O. Box 218, Yorktown Heights, NY 10598,
bge@watson.ibm.com} }
\author{Thomas Ferguson\affil{Vassar College, Dept. of Physics \& Astronomy, Box 745,
Poughkeepsie, NY 12604; thferguson@vassar.edu }}
\author{Brendan Mullan\affil{Vassar College, Dept. of Physics \& Astronomy, Box 745,
Poughkeepsie, NY 12604 and Colgate University, Dept. of Astronomy, Hamilton, NY;
bmullan@mail.colgate.edu }}
\begin{abstract}
GEMS and GOODS fields were examined to $z\sim$1.4 for galaxy
interactions and mergers. The basic morphologies are familiar:
antennae with long tidal tails, tidal dwarfs, and merged cores;
M51-type galaxies with disk spirals and tidal arm companions;
early-type galaxies with diffuse plumes; equal-mass
grazing-collisions; and thick J-shaped tails beaded with star
formation and double cores. One type is not common locally and is
apparently a loose assemblage of smaller galaxies. Photometric
measurements were made of the tails and clumps, and physical sizes
were determined assuming photometric redshifts. Antennae tails are a
factor of $\sim3$ smaller in GEMS and GOODS systems compared to
local antennae; their disks are a factor of $\sim2$ smaller than
locally. Collisions among early type galaxies generally show no fine
structure in their tails, indicating that stellar debris is usually
not unstable. One exception has a $5\times10^9$ M$_\odot$ smooth red
clump that could be a pure stellar condensation. Most tidal dwarfs
are blue and probably form by gravitational instabilities in the
gas. One tidal dwarf looks like it existed previously and was
incorporated into the arm tip by tidal forces. The star-forming
regions in tidal arms are $10$ to $1000$ times more massive than
star complexes in local galaxies, although their separations are
about the same. If they all form by gravitational instabilities,
then the gaseous velocity dispersions in interacting galaxies have
to be larger than in local galaxies by a factor of $\sim5$ or more;
the gas column densities have to be larger by the square of this
factor.
\end{abstract}


\section{Introduction}

Galaxy interactions and mergers are observed at all redshifts and
play a key role in galaxy evolution. Two percent of local galaxies
are interacting or merging (Athanassoula \& Bosma 1985; Patton et
al. 1997), and this fraction is larger at high redshift (e.g.,
Abraham et al. 1996b; Neuschaefer et al. 1997 ; Conselice et al.
2003; Lavery et al. 2004; Straughn et al. 2006; Lotz et al. 2006,
and others).  Conselice (2006a) estimates that massive galaxies have
undergone about 4 major mergers by redshift 1. Toomre (1977)
described a sequence of merger activity ranging from separated
galaxies with tails and a bridge between them, to double nuclei in a
common envelope with tails, to merged nuclei with tails.
Ground-based (Hibbard \& van Gorkom 1996) and space-based (Laine et
al. 2003; Smith et al. 2007) observations of this sequence show
optical, infrared, and radio activity in the tails and nuclei.

High resolution images and numerical simulations of nearby
interactions demonstrate how star formation and morphology are
affected. General reviews of interaction simulations are given by
Barnes \& Hernquist (1992) and Struck (1999). The initial galaxy
properties, such as mass, rotational velocity, gas content and dark
matter content, and their initial separations and velocity vectors,
all play a role in generating structure. The viewing angle also
affects the morphology. Early-type galaxies with little gas are
expected to display smooth plumes and shells, while spiral
interactions and mergers should exhibit clumpy star formation along
tidal tails, and condensations of material at the tail ends. Equal
mass companions may show bridges between them. A prominent example
of a tidal interaction is the Antennae (NGC4038/9), a merging pair
of disk galaxies with rampant star formation in the central regions,
including young globular clusters (Whitmore et al. 2005). Its
interaction was first modeled by Toomre \& Toomre (1972). The
Cartwheel galaxy is a collisional ring system rimmed with star
formation from a head-on collision (Struck et al. 1996). Sometimes
polar-ring or spindle galaxies are the result of perpendicular
collisions (Struck 1999). The Mice (NGC 4676) has a long narrow
straight tail and a curved tidal arm (Vorontsov-Velyaminov 1957;
Burbidge \& Burbidge 1959); numerical simulations reproduce both
features well in a model with a halo:(disk+bulge) mass ratio of 5
(Barnes 2004).   The Superantennae (IRAS 19254-7245) is a pair of
infrared-luminous merging giant galaxies having Seyfert and
starburst nuclei and $\sim200$ kpc tails with a tidal tail dwarf
(Mirabel, Lutz, \& Maza, 1991). The Leo Triplet includes NGC 3628
with an 80 kpc stellar tail containing star-forming complexes with
masses up to $10^6$ M$_{\odot}$ (Chromey et al. 1998). The Tadpole
galaxy UGC10214 (Tran et al. 2003; de Grijs et al. 2003; Jarrett et
al. 2006), the IC2163/NGC2207 pair (Elmegreen et al. 2001, 2006),
and Arp 107 (Smith et al. 2005) are all interacting systems observed
with HST and SST and modeled in simulations. Many local mergers have
intense nuclear activity, such as the Seyfert galaxy NGC 5548, which
also has an 80 kpc long, low surface brightness (V=27-28 mag
arcsec$^{-2}$) tidal tail and a 1-arm diffuse spiral (Tyson et al.
1998).

The GEMS (Galaxy Evolution from Morphology and SEDs; Rix et al.
2004), GOODS (Great Observatories Origins Deep Survey; Giavalisco et
al. 2004), and UDF (Ultra Deep Field; Beckwith et al. 2006) surveys
done with the HST ACS (Hubble Space Telescope Advanced Camera for
Surveys) have enabled high resolution studies of the morphology of
intermediate and high redshift galaxies. Light distribution
parameters such as the Gini coefficient (Lotz et al. 2006) and
concentration index, asymmetry, and clumpiness (CAS; Conselice 2006)
have been applied to galaxies in these fields to study possible
merger systems. For GEMS and GOODS, John Caldwell of the GEMS team
has posted images (archive.stsci.edu/prepds/gems/datalist.html) of
several galaxies from each field, including peculiar and interacting
systems with tails and bridges. Here we examine the entire GEMS and
GOODS fields systematically for such galaxies and study their tails,
bridges, and star-forming regions. Their properties are useful for
understanding interactions and interaction-triggered star formation,
and for probing the relative dark matter content (e.g., Dubinski,
Mihos, \& Hernquist 1999).

\section{The Sample of Interactions and Mergers}
\label{sect:sample}

The GOODS and GEMS images from the public archive were used for this
study. They include exposures in 4 filters for GOODS:  F435W
(B$_{435}$), F606W (V$_{606}$), F775W (i$_{775}$), and F850LP
(z$_{850}$); and 2 filters (V$_{606}$ and z$_{850}$) for GEMS. The
public images were drizzled to produce final archival images with a
scale of 0.03 arcsec per px. GEMS, which incorporates the southern
GOODS survey (Chandra Deep Field South, CDF-S)  in the central
quarter of its field, covers 28 arcmin x 28 arcmin; there are 63
GEMS and 18 GOODS images that make up the whole field.  The GOODS
images have a limiting AB mag of V$_{606}$= 27.5 for an extended
object, or about two mags fainter than the GEMS images. There are
over 25,000 galaxies catalogued in the COMBO-17 survey (Classifying
Objects by Medium-Band Observations, a spectrophotometric 17-filter
survey; Wolf et al. 2003), and 8565 that are cross-correlated with
the GEMS survey (Caldwell et al. 2005).

\begin{figure}\epsscale{0.8} \plotone{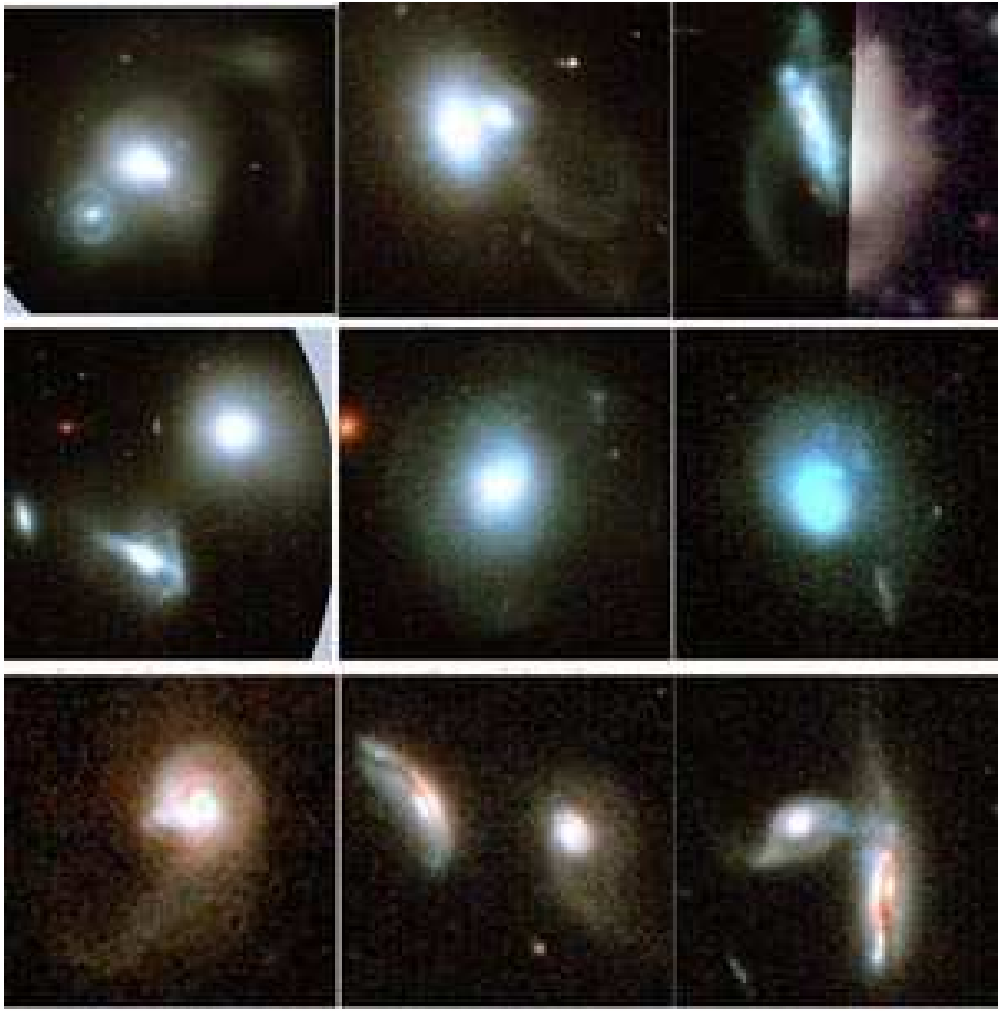}\caption{Color
images of galaxies in the GEMS and GOODS fields with smooth diffuse
tidal debris. The galaxy at the top right, number 3 in Table 1, is
only partially covered by the GEMS field; the right-hand portion of
the image is from ground-based observations. The smooth debris is
presumably from old stars that were spread out during the
interaction. A few small star-formation patches are evident in some
cases. The clump in the upper right corner of the galaxy 1 image
could be a rare example of a gravitationally driven condensation in
a pure-stellar arm. The smooth arcs and spirals in this and other
images are probably a combination of orbital debris and flung-out
tidal tails. The galaxy numbers, as listed in Table 1, are 1 through
9, as plotted from left to right and top to bottom. (Image quality
degraded for astroph.)}\label{fig1diffuse}\end{figure}

\begin{figure}\epsscale{0.8}
\plotone{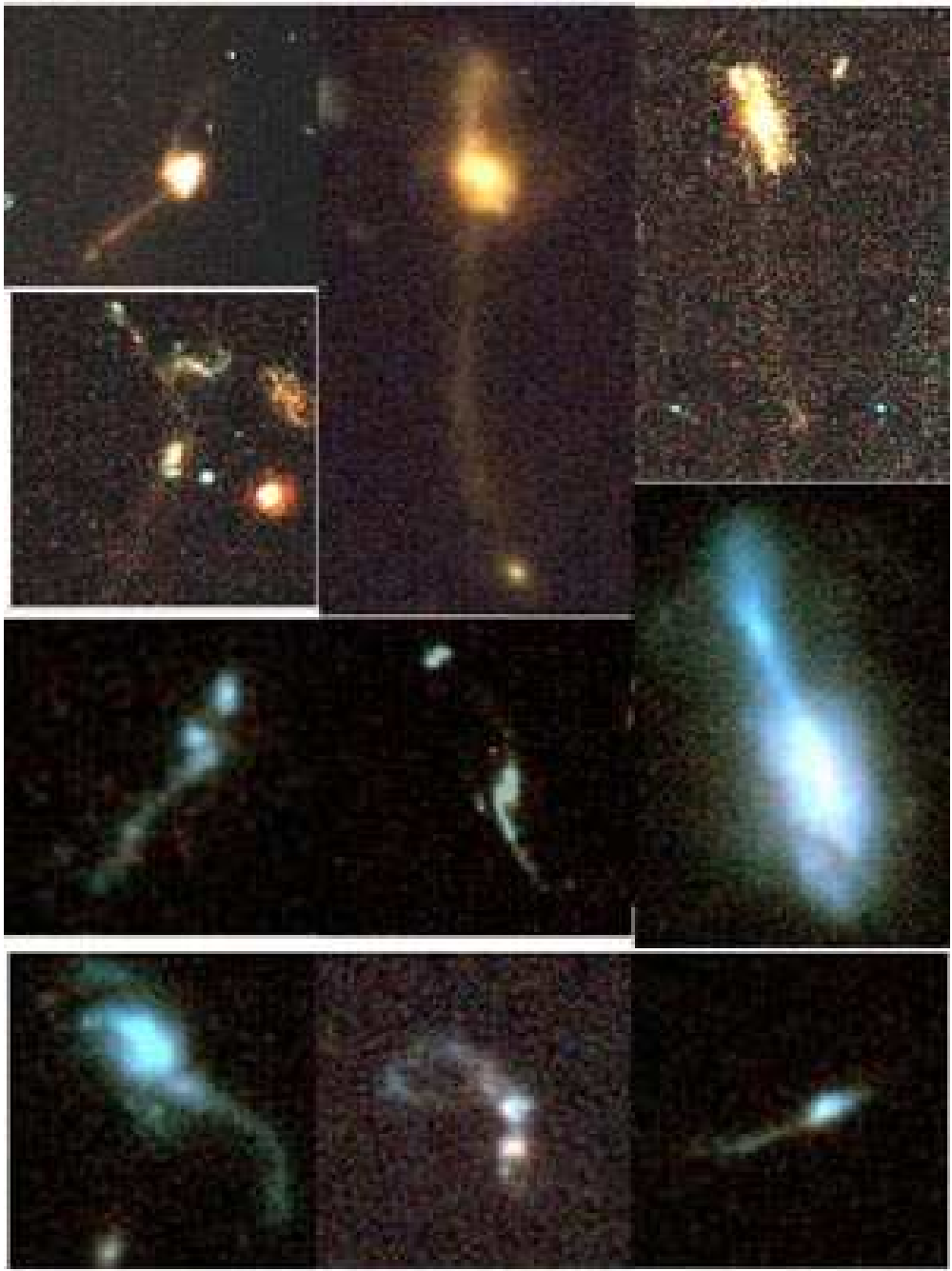}\caption{Color images of interacting
antennae galaxies with long and structured tidal arms. Galaxy
numbers, in order, are 15 through 24. Several have dwarf galaxy-like
condensations at the arm tips or broad condensations midway out in
the arms. The dwarf elliptical at the tip of the tidal arm in galaxy
20 might have existed before the interaction and been placed there
by tidal forces; the main body of this system has a double nucleus
from the main interaction. (Image quality degraded for astroph.)
}\label{fig2bants}\end{figure}

\begin{figure}\epsscale{0.8}
\plotone{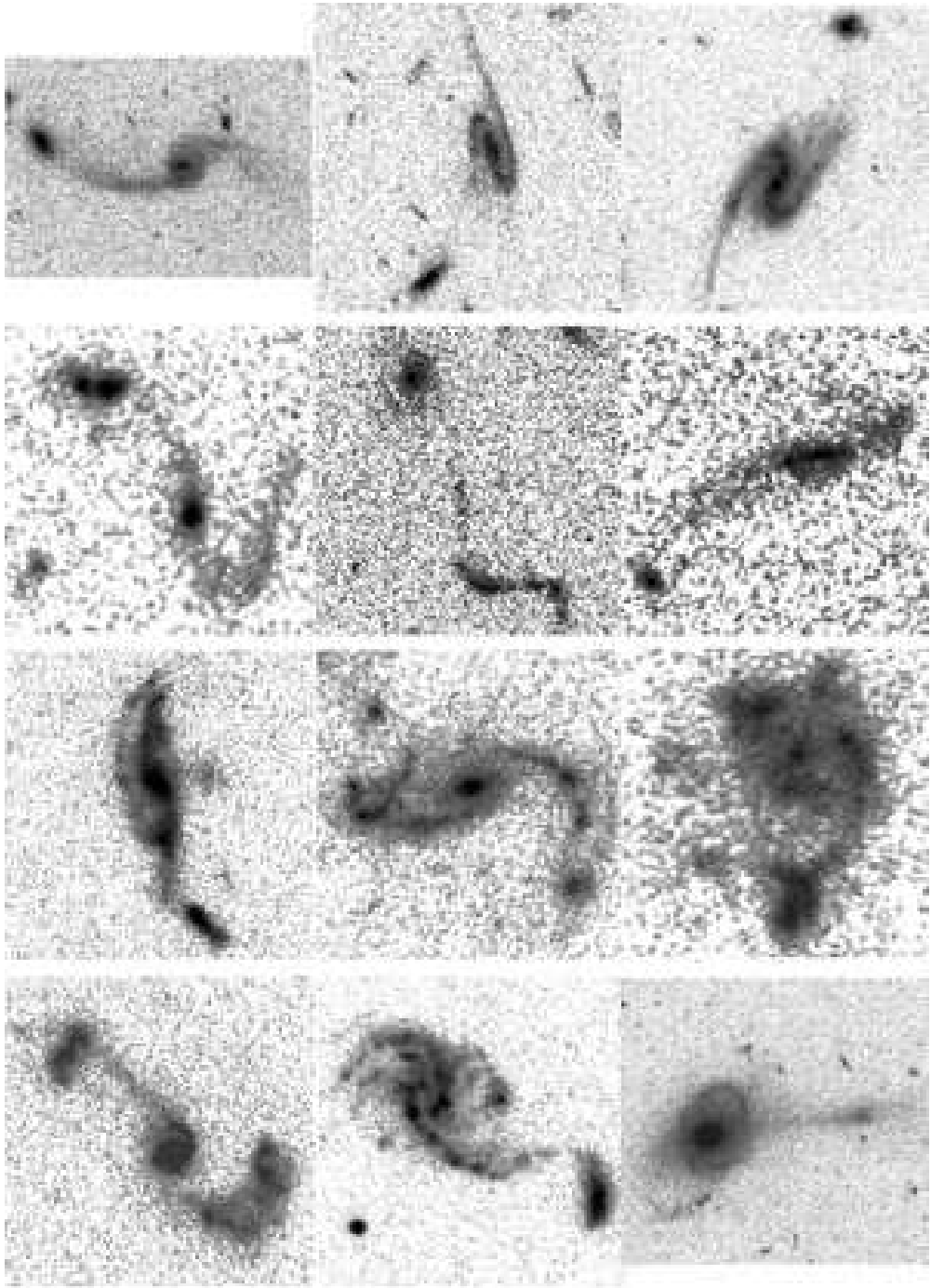}\caption{M51-type galaxies are shown as
logarithmic grayscale V-band images. In order, the galaxy numbers
are 33 through 44. The linear streak in galaxy 44 could be orbital
debris from the small companion on the right. (Image quality
degraded for astroph.) }\label{fig3M51types}\end{figure}

\begin{figure}\epsscale{.9} \plotone{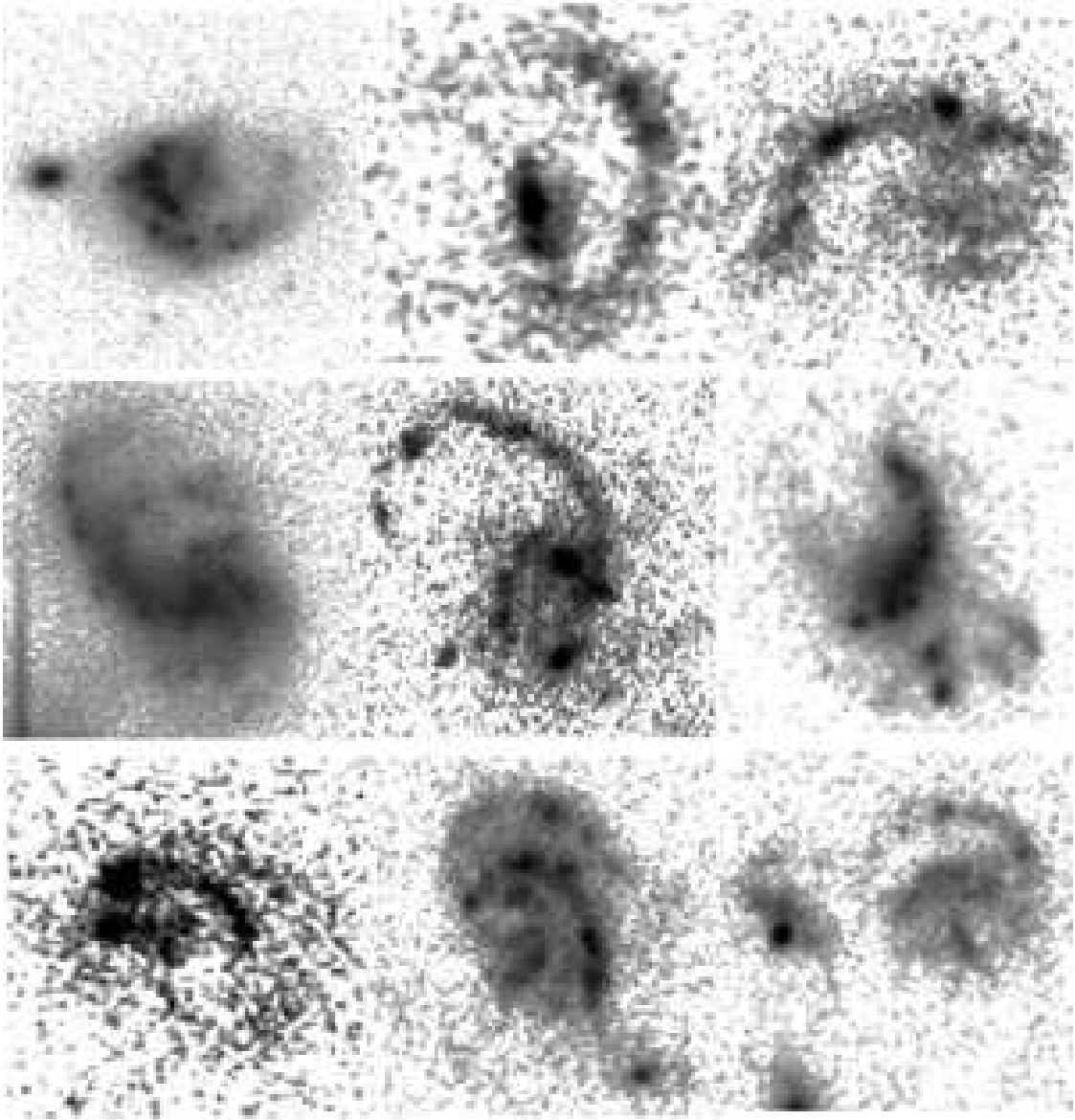}
\caption{Shrimp galaxies, named because of their curved tails, are
shown as logarithmic V-band images. In order, they are numbers 55
through 63. (Image quality degraded for
astroph.)}\label{fig4shrimps}\end{figure}

\begin{figure}\epsscale{1.0}
\plotone{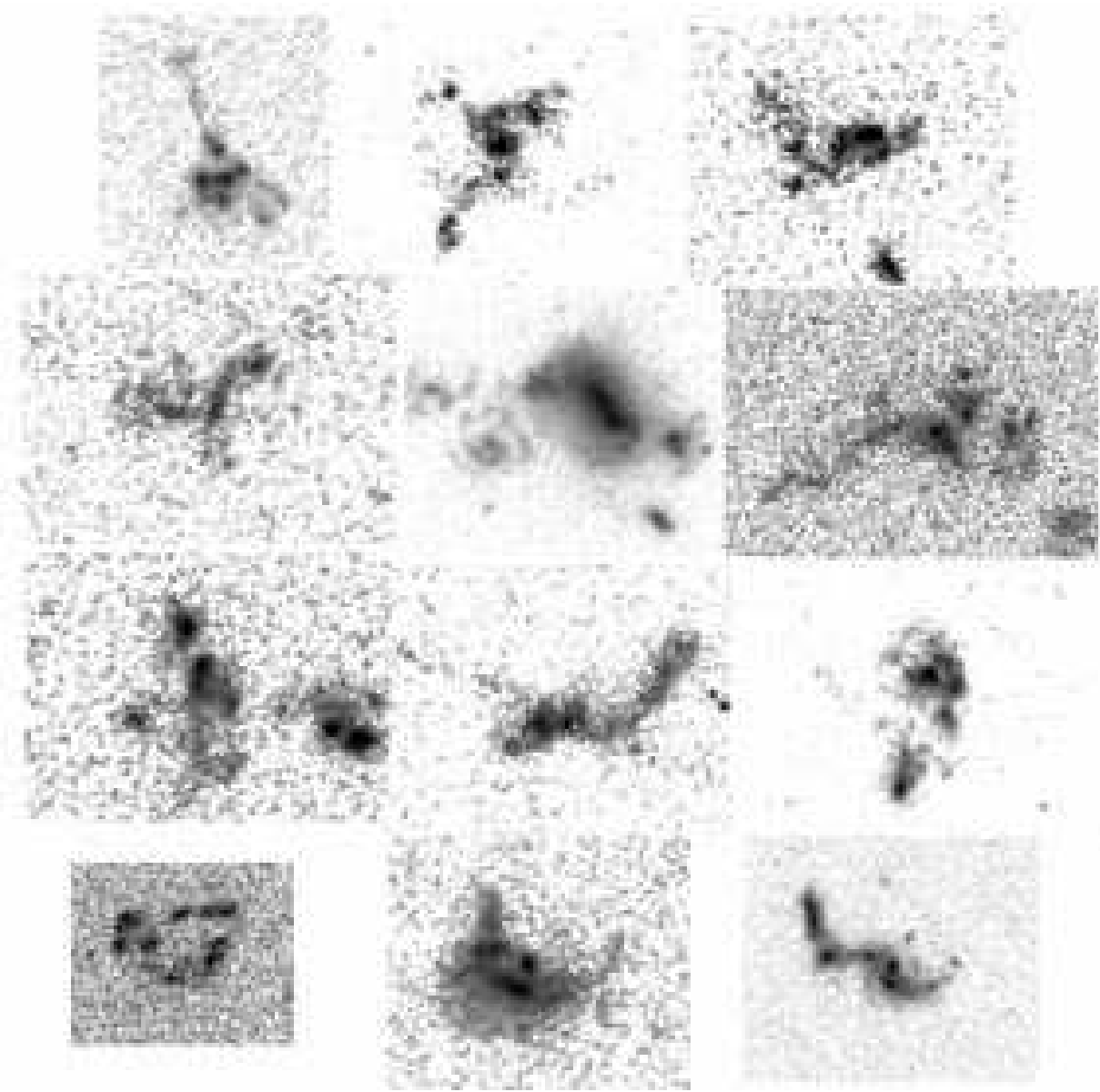}\caption{Assembly galaxies look like they
are being assembled through mergers. In order: galaxy 74 through 85.
}\label{fig:wrecks}\end{figure}

\begin{figure}\epsscale{1.0}
\plotone{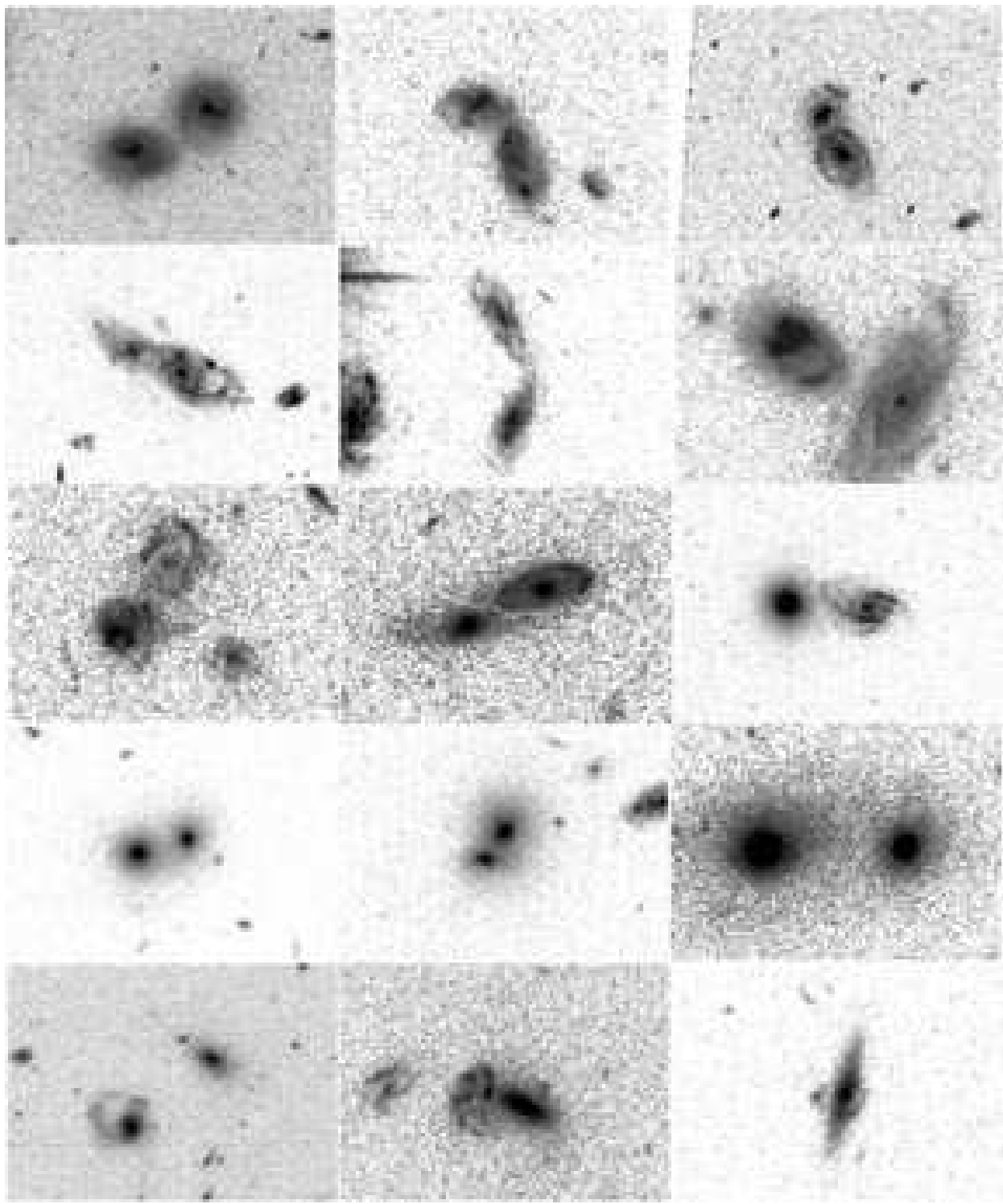}\caption{Galaxies with approximately
equal-mass grazing companions, in order, are 86 through 100.
}\label{fig:equals}\end{figure}

Interacting galaxies with tails, bridges, diffuse plumes and other
features were identified by eye on the online Skywalker images and
examined on high resolution $V_{606}$ fits images. The lower limit
to the length of detectable tails is about 20 pixels. Snapshots of
several different morphologies for interacting galaxies are shown in
Figures \ref{fig1diffuse}-\ref{fig:equals}. Out of an initial list
of about 300 galaxies, a total of 100 best cases are included in our
sample: 14 diffuse types, 18 antennae types, 22 M51 types, 19 shrimp
types, 15 equal mass interactions, and 12 assemblies, as we describe
below.

GEMS and GOODS galaxy redshifts were obtained from the COMBO-17 list
(Wolf et al. 2003). Our sample ranges from redshift $z=0.1$ to 1.4
in an area of $2.8\times10^6$ square arcsec. The linear diameters of
the central objects were determined from their angular diameters and
redshifts using the appropriate conversion for a $\Lambda$CDM
cosmology (Carroll et al., 1992; Spergel et al., 2003). The range is
$\sim3$ to 33 kpc. Projected tail lengths were measured in a
straight line from the galaxy center to the 2$\sigma$ noise limit
(25.0 mag arcsec$^{-2}$) in the outer tail.

Photometry was done on the whole galaxies, on each prominent
star-forming clump, and on the tails using the IRAF task {\it
imexam}. A box of variable size was defined around each feature; the
outer limits of the boxes were chosen to be where the clump
brightness is about 3 times the surrounding region. Sky subtraction
was not done because the background is negligible. The photometric
errors are $\sim$0.1-0.2 mag for individual clumps. The V$_{606}$
surface brightnesses of the tidal tails were determined using {\it
imagej} (Rasband 1997) to trace freehand contours around the tails,
so that they could be better defined than with rectangular or
circular apertures.

Figure \ref{fig1diffuse} shows galaxies with diffuse plumes and
either no blue star formation patches or only a few tiny patches
(e.g., galaxies number 5 and 6); we refer to these interactions as
diffuse types. The colors of the plumes match the colors of the
outer parts of the central galaxies, indicating the plumes are
tidally shorn stars with little gas. There is structure in most of
the plumes consisting of arcs or sharp edges. This is presumably
tidal debris from early type galaxies with little or no gas (e.g.
Larson \& Tinsley 1978; Malin \& Carter 1980; Schombert et al.
1990). This type of interaction is relatively rare in the GEMS and
GOODS images, perhaps because the tidal debris is faint. The best
cases are shown here and they all have relatively small redshifts
compared to the other interaction types (the average $z$ is 0.23 and
the maximum $z$ is 0.69).

The image in the top left panel of Figure \ref{fig1diffuse} (galaxy
1) has a giant diffuse clump in the upper right corner. This could
be a condensation in the tidal arm, or it could be another galaxy.
In either case, it has the same color as the rest of the tidal arm
nearby. That is, $V_{606}-z_{850} = 0.90\pm0.5$ for the clump and
also in six places along the tail; the color is essentially the
same, $0.94\pm0.05$, in the core of the galaxy. The absolute
magnitude of the clump is $M_V=-18.41$ for redshift $z=0.15$. The
mass is $\sim5\times10^9$ M$_\odot$ (Sect. \ref{sect:clump}). If
this clump is a condensation in the tail, then it could be a rare
case where a pure stellar arc has collapsed gravitationally into a
gas-free tidal dwarf. The final result could be a dwarf elliptical.
Usually tidal dwarfs form by gaseous condensations in tidal arms
(Wetzstein, Naab, \& Burkert 2007).

Figure \ref{fig2bants} shows interactions that resemble the local
Antennae pair, so we refer to them as antennae types. These types
have long tidal tails and double nuclei or highly distorted centers
that appear to be mergers of disk galaxies. Note that antennae are
not the same as ``tadpole'' galaxies (Elmegreen et al. 2005a; de
Mello et al. 2006; Straughn et al. 2006), which have one main clump
and a sometimes wiggly tail that may contain smaller clumps. Some
antennae have giant clumps near the ends of the tails which could
have formed there (galaxies 16 and 17) and are analogous to the
clump at the end of the Superantennae (Mirabel et al. 1991). Galaxy
18 is in a crowded field with at least two long tidal arms; here we
consider only the tail system in the north, which is in the upper
part of the figure. These long-tail systems are relatively rare and
all the best cases are shown in the figure; their average redshift,
0.70, is typical for GEMS and GOODS fields. Galaxy 24 is somewhat
like a tadpole galaxy, but its very narrow tail and protrusion on
the anti-tail side of the main clump are unlike structures seen in
tadpoles of the Ultra Deep Field.

For the antennae galaxies in Figure \ref{fig2bants}, the tails have
an average (V-z) color that is negligibly bluer, 0.10$\pm$0.25 mag,
than the central disks. In a study of tidal features in local Arp
atlas galaxies, Schombert et al. (1990) also found that the tail
colors are uniform and similar to those of the outer disks. They
noted that the most sharply-defined tails are with spiral systems
and the diffuses plumes are with ellipticals. This correlation may
be true here also, but it is difficult to tell from Figure
\ref{fig1diffuse} whether the smooth distorted systems are
intrinsically disk-like.

Galaxy 20 in Figure \ref{fig2bants} is an interesting case. It has
an elliptical clump at the end of its tail that could be one of the
collision partners. There are two central galaxy cores, however, and
their interaction may have formed the tidal arms without this
companion. Furthermore, the clump at the tip is aligned
perpendicular to the tail, which is unusual for a tidal dwarf.  Thus
it is possible that the clump was a pre-existing galaxy lying in the
orbital plane of one of the larger galaxies now at the center.
Presumably this former host is the galaxy currently connected to the
dwarf by the tidal arm. The interaction could have swung it around
to its current position at the tip. A similar case occurs for the
local IC 2163/NGC 2207 pair, which has a spheroidal dwarf galaxy at
the tip of its tidal arm (Elmegreen et al. 2001). Such swing-around
dwarfs should have the same dynamical origin as the large pools of
gas and star formation that are at the tips of superantenna-type
galaxies; i.e. the whole outer disk moves to this position during
the interaction (Elmegreen et al. 1993; Duc et al. 1997).

Figure \ref{fig3M51types} shows examples of interactions that we
refer to as M51-type galaxies, where the tidal arms can be bridges
that connect the main disk galaxy to the companion (galaxy 33), or tails
on the opposite side of the companion (e.g., galaxies 34 and 35), or
both (galaxy 36). In galaxy 44, the tidal arm looks like the debris
path of a pre-existing galaxy that lies at the right; the orbit path
apparently curves around on the left.  The M51-types usually have
strong spirals in the main disk. In the top row, the tails and
bridges are thin and diffuse. The galaxy on the left in the lower
row (galaxy 42) has a thick, fan-shaped tail opposite the companion.
Some bridges have star formation clumps (galaxy 40) and others
appear smooth (galaxy 33). Interactions like this, especially those
with small companions, are more common than the previous two types
and only a few best cases are shown in Figure \ref{fig3M51types} and
discussed in the rest of this paper.

Figure \ref{fig4shrimps} shows examples of galaxies dominated by one
highly curved, dominant arm and large, regularly-spaced clumps of
star formation. We call these ``shrimp'' galaxies because of their
resemblance to the tail of a shrimp. Although their star formation
indicates they contain gas and therefore are disk systems, there are
no well-defined spirals (except for the prominent arm), merging
cores, or obvious central nuclei. The clumps resemble the
beads-on-a-string star formation in spiral density waves and
probably have the same origin, a gravitational instability
(Elmegreen \& Elmegreen 1983; Kim \& Ostriker 2006; Bournaud, Duc,
\& Masset 2003). The J-shaped morphology is reminiscent of the 90
kpc gas tail of M51 (Rots et al. 1990) and the 48 kpc gas tail
observed in NGC 2535 (Kaufman et al. 1997). Rots el al. point out
that the M51 gas tail is much broader (10 kpc) than the narrow tails
seen in merging systems like the Antennae. The broad tail in galaxy
42 (Fig. 3) is similar to the M51 tail. Sometimes there is a bright
tail with no obvious companion (galaxies 56, 57, and 60); one of
these, galaxy 56, was in our ring galaxy study (Elmegreen \&
Elmegreen 2006). Asymmetric, strong arm galaxies like this are not
common in GEMS and GOODS; this figure shows the best cases.

Figure \ref{fig:wrecks} has a selection of irregular galaxies that
appear to be interactions. Most of them suggest an assembly of small
pieces, so we refer to them as assembly types. If they were slightly
more round in overall shape, with more obvious interclump emission,
then we would classify them as clump-clusters, as we did in the UDF
(Elmegreen et al. 2005a). The galaxy in the lower left (galaxy 83) is like
this.  The resemblance of these types to clump-clusters suggests that some of the
clumps are accreted from outside the disk and others form
from gravitational instabilities in a pre-existing gas disk, as
suggested previously (Elmegreen \& Elmegreen 2005). The system in the lower right (galaxy 85)
could be interacting spirals, or a triple system, or a bent chain
(as studied in Elmegreen \& Elmegreen 2006). There are many examples of
highly irregular galaxies like these in the GEMS and GOODS fields;
indeed most galaxies at $z>1.5$ are peculiar in this sense
(Conselice 2005). In what follows, we discuss only these 12 galaxies.

Figure \ref{fig:equals} has samples of grazing or close
interactions, with spirals at the top of the page (numbers 86-93),
ellipticals lower down (numbers 95-97) and two polar-ring galaxies
(numbers 99 and 100) in the lowest row at the middle and right
bottom. We refer to these paired systems as ``equals'' because their
distinguishing feature is that the two galaxies have comparable
size. The pair number 89 has a bright oval in the smaller galaxy,
which is characteristic of recent tidal forces for an in-plane,
prograde encounter such as IC2163/NGC2207 (Sundin 1993; Elmegreen et
al. 1995 ). There is a spiral-elliptical pair on the right in the
middle row (galaxy 94). Double ellipticals in the UDF were studied
previously (Elmegreen et al. 2005a, Coe et al. 2006). Near neighbors
like this have been studied previously in the GEMS field; 6 double
systems out of 379 red sequence galaxies were identified as being
dry merger candidates, as reproduced in simulations (Bell et al.
2006). The models of mergers of early-type systems by Naab et al.
(2006) apparently account for kinematic and isophotal properties of
ellipticals better than the formation of ellipticals through
late-type mergers alone. For the pairs in our figure, both
components have the same COMBO17 redshift. There are many other
examples of close galaxy groups and near interactions in the GEMS
and GOODS surveys. In what follows we discuss only the properties of
those shown in Figure \ref{fig:equals}.

The interacting types shown in the figures are meant to be as
distinct as possible.  These and other good cases are listed in
Table 1 by running number, along with their COMBO-17 catalog number,
redshift, and R magnitude. There is occasionally some ambiguity and
overlap in the interaction types, particularly between M51-types and
shrimps when the M51-types have small or uncertain companions at the
ends of their prominent tails. Projection effects can lead to
uncertainties in the classifications as well, particularly for
antennae whose tails may be foreshortened. Nevertheless, these
divisions serve as a useful attempt to sort out the most prominent
features among interacting galaxies. There are numerous other
galaxies in GEMS and GOODS that are apparently interacting, but most
of them are too highly distorted to indicate the particular physical
properties of interest here, namely, disk-to-halo mass ratio and
star formation scale.

\section{Photometric Results}

\subsection{Global galaxy properties}

\begin{figure}\epsscale{1.0}
\plotone{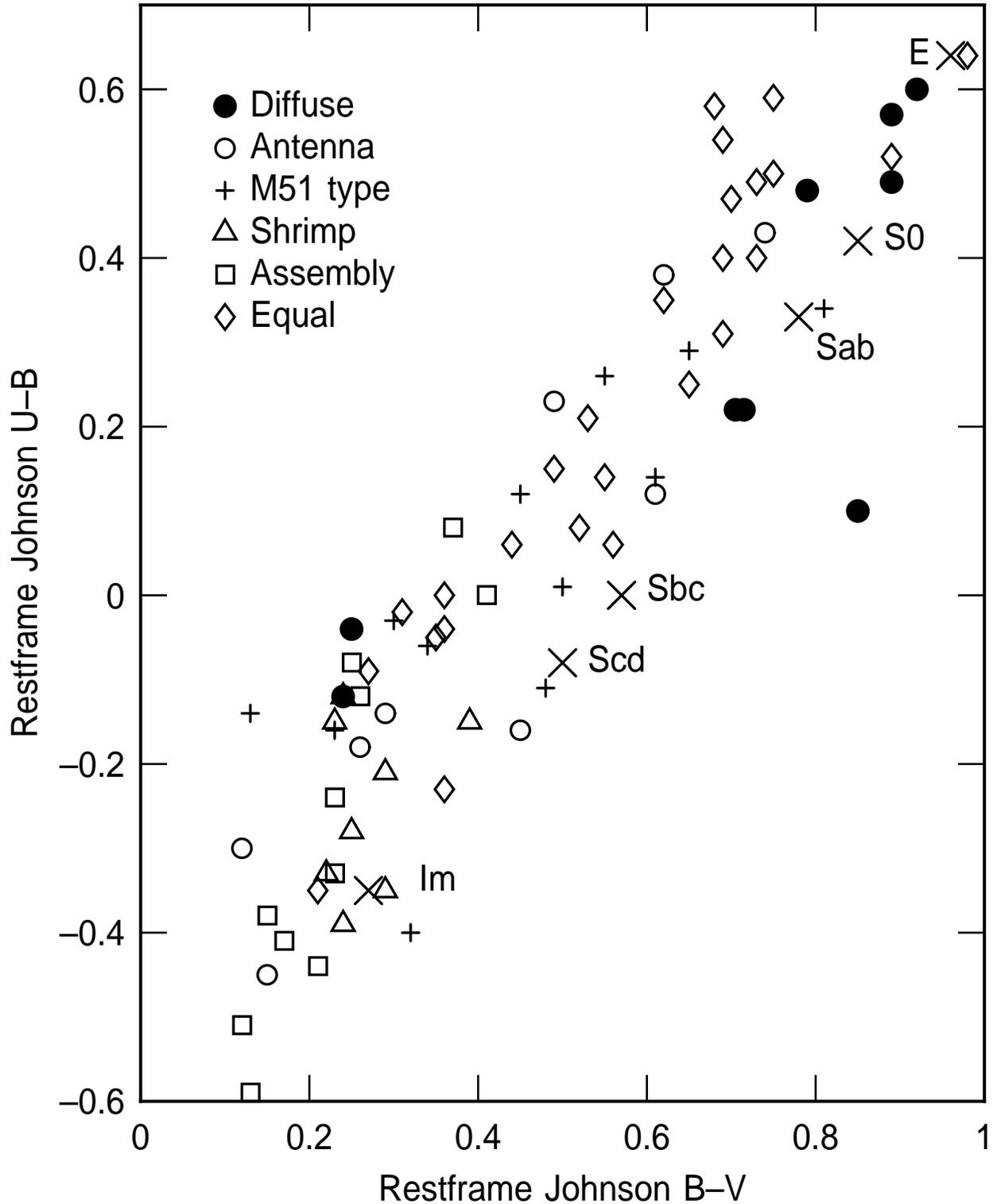}\caption{Restframe (U-B) and (B-V) integrated colors
for interacting galaxies in the GEMS and GOODS fields, from
COMBO-17. The reddest tend to be the diffuse types, which are
presumably dry mergers, and the bluest are the assembly types, which
could be young proto-galaxies. Crosses indicate standard Hubble types, measured by Fukugita et al. (1995).}\label{fig:shrimf6}\end{figure}

The integrated Johnson restframe (U-B) and (B-V) colors from
COMBO-17 for the observed galaxies with measured redshifts are
shown in a color-color diagram in Figure \ref{fig:shrimf6}. The
crosses in the diagram are Johnson colors for standard Hubble types
(Fukugita et al. 1995). Our sample of galaxies spans the range of
colors from early to late Hubble types, although the bluest are
bluer than standard irregular galaxies (a typical Im has
U-B$=-0.35$, B-V$=0.27$).  The reddest galaxies tend to be the
diffuse types, thought to originate with ellipticals involved in
interactions. The two reddest galaxies in our sample are the diffuse
types number 1 and 2 in Figure \ref{fig1diffuse}. The bluest tend to
be the assemblies, consistent with their having formed recently.

\begin{figure}\epsscale{1.0}
\plotone{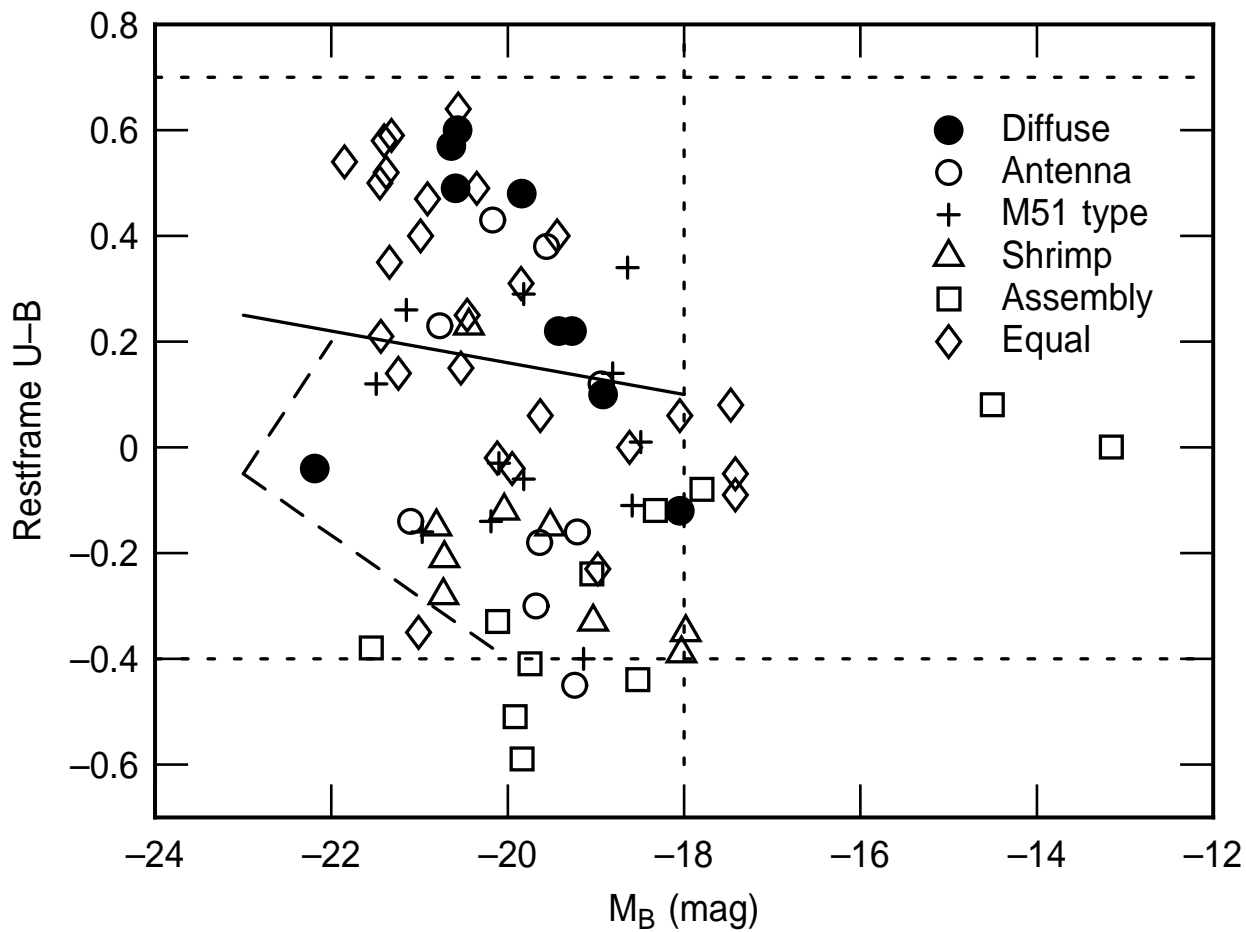}\caption{Restframe Johnson U-B integrated color
versus absolute restframe $M_B$, from COMBO-17. The solid line
separates the red sequence and blue cloud (Conselice 2006b). Color limits for local
galaxies are indicated by the horizontal short-dashed lines; local galaxies are brighter than the vertical line. The local blue cloud galaxies are approximately delimited on the left side of the diagram by the long-dashed lines. Thus, most of our observed galaxies fall near the local galaxy colors and magnitudes.}
\label{fig:shrimf15}\end{figure}

Figure \ref{fig:shrimf15} shows a restframe color-magnitude diagram.
Early and late type galaxies usually separate into a ``red
sequence'' and a ``blue cloud'' on such a diagram (Baldry et al.
2004; Faber et al. 2005).  The solid line indicates the boundary
between these two regions from a study of 22,000 nearby galaxies
(Conselice 2006b). The short-dashed lines are the limits of the
Conselice (2006) survey; local galaxies are brighter than the
vertical short-dashed line and their colors lie between the
horizontal short-dashed lines. The long-dashed lines approximately
outline the bright limit for the local blue cloud galaxies. Our
galaxies fall in both the red sequence and the blue cloud. The
restframe colors in Figure \ref{fig:shrimf15} are consistent with
their morphological appearances. The red sequence galaxies in the
figure usually appear smooth (the diffuse types) or lack obvious
huge star formation clumps (the equal mass mergers), while the blue
cloud galaxies usually have patches of star formation (the
M51-types, shrimps, assemblies, and many antennae). We see now why
the redshifts of the diffuse galaxies ($z<0.3)$ are much lower than
the others: this is a selection effect for the ACS camera. These
tails comprise old stellar populations without star-forming clumps,
and their intrinsic redness makes them difficult to see at high
redshifts. Also, they tend to have intrinsically low surface
brightnesses because of a lack of star formation, and cosmological
dimming makes them too faint to see at high redshift. Hibbard \&
Vacca (1997) note that it is difficult to detect tidal arms beyond
$z\sim1.5$.

\subsection{Clump properties}
\label{sect:clump}

Prominent star-forming clumps are apparent in many of the
interacting galaxies.  Their sizes and magnitudes were measured
using rectangular apertures. The observed magnitudes were converted
to restframe B magnitudes whenever possible, using linear
interpolations between the ACS bands. For example, GEMS observations
are at two filters, $V_{606}$ and $z_{850}$. GEMS galaxies with
redshifts $z$ between 0.39 ($=606/435-1$) and 0.95 ($=850/435-1$)
were assumed to have restframe blue luminosities given by
$L_{B,rest}= L_{V,obs}(0.95-z)/(0.95-0.39) + L_{z,obs}
(z-0.39)/(0.95-0.39)$. The restframe B magnitude is then $-2.5 \log
L_{B,rest}$. For GOODS galaxies, the conversions were divided into 3
redshift bins to make use of the 4 available filters, and a linear
interpolation was again applied to get restframe clump magnitudes.
For the GOODS galaxies, the restframe magnitudes determined by
interpolation between the nearest 2 filters among the 4 filters are
within $\pm$0.2 mag of the restframe magnitudes determined from only
the V and z filters. Thus, the GEMS interpolations are accurate to
this level. (We do not include corrections for intergalactic
absorption in these colors, because we are comparing them directly
with their parent galaxy properties. Below, when we convert the
colors and magnitudes to masses and ages, absorption corrections are
taken into account.)

\begin{figure}\epsscale{1.0} \plotone{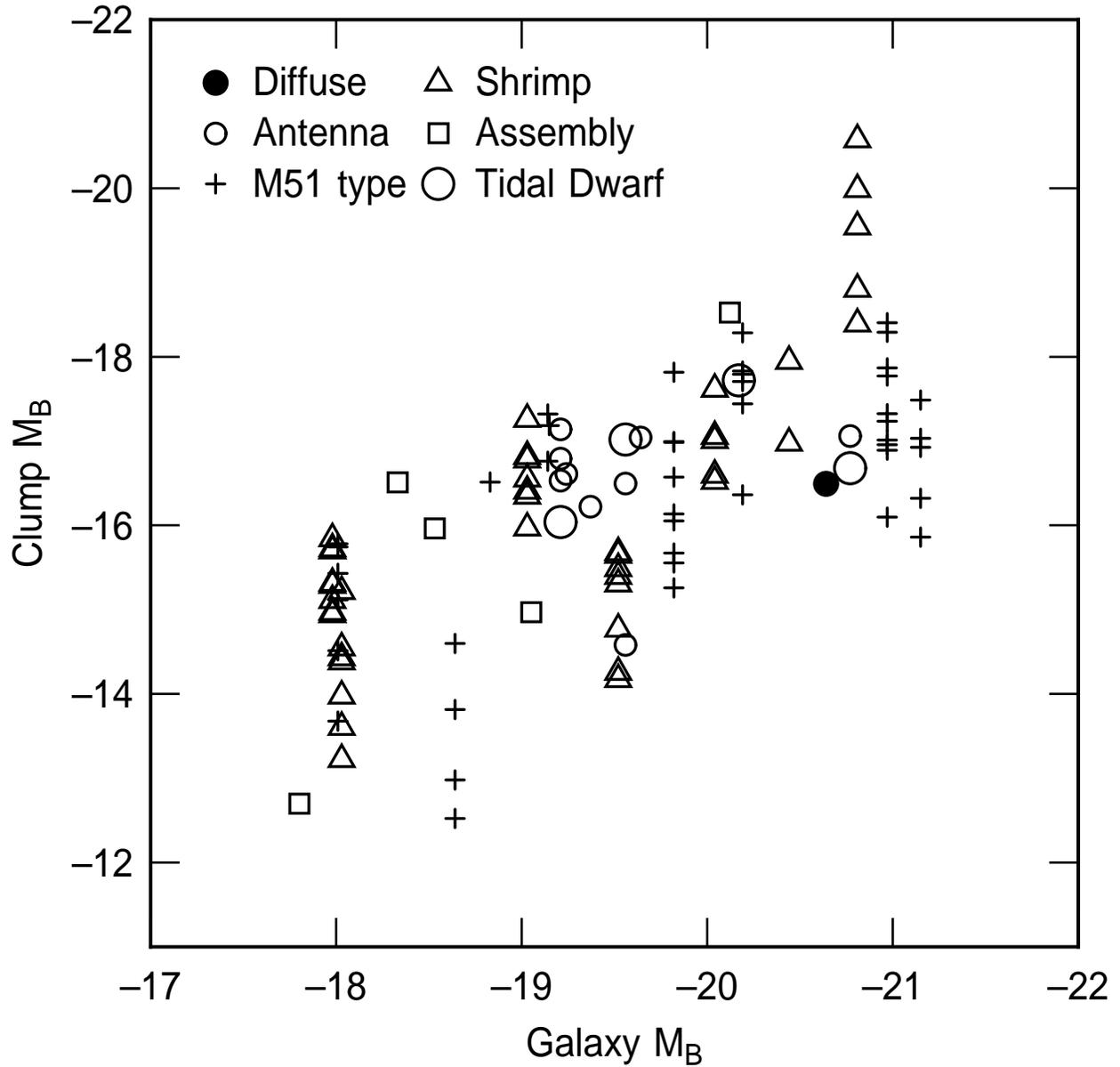}\caption{Restframe
B absolute magnitudes of star-forming clumps versus integrated
galaxy restframe magnitudes. The correlation is also found for local
galaxies.}\label{fig:shrimf9}\end{figure}

The apparent restframe B magnitudes of the clumps were converted to
absolute restframe B magnitudes using photometric redshifts and the
distance modulus for a $\Lambda$CDM cosmology. These absolute clump
magnitudes are shown as a function of absolute galaxy magnitude in
Figure \ref{fig:shrimf9}. The clump absolute B magnitudes scale
linearly with the galaxy magnitudes. The clumps are typically a kpc
in size ($\sim3$ to 8 pixels across), comparable to star-forming
complexes in local galaxies (Efremov 1995), which also scale with
galaxy magnitude (Elmegreen et al. 1996; Elmegreen \& Salzer 1999).

Clump ages and masses were estimated by comparing observed clump
colors, magnitudes, and redshifts with evolutionary models that
account for bandshifting and intergalactic absorption and that
assume an exponential star formation rate decay (see Elmegreen \&
Elmegreen 2005). Internal dust extinction as a function of redshift
is taken from Rowan-Robinson (2003).  The GEMS galaxy clumps only
have (V$_{606}$-z$_{850}$) colors, so the ages are not well
constrained. For the GOODS galaxies, the additional B and I filters
help place better limits on the ages, although there is still a wide
range of possible fits.

\begin{figure}\epsscale{1.0}
\plotone{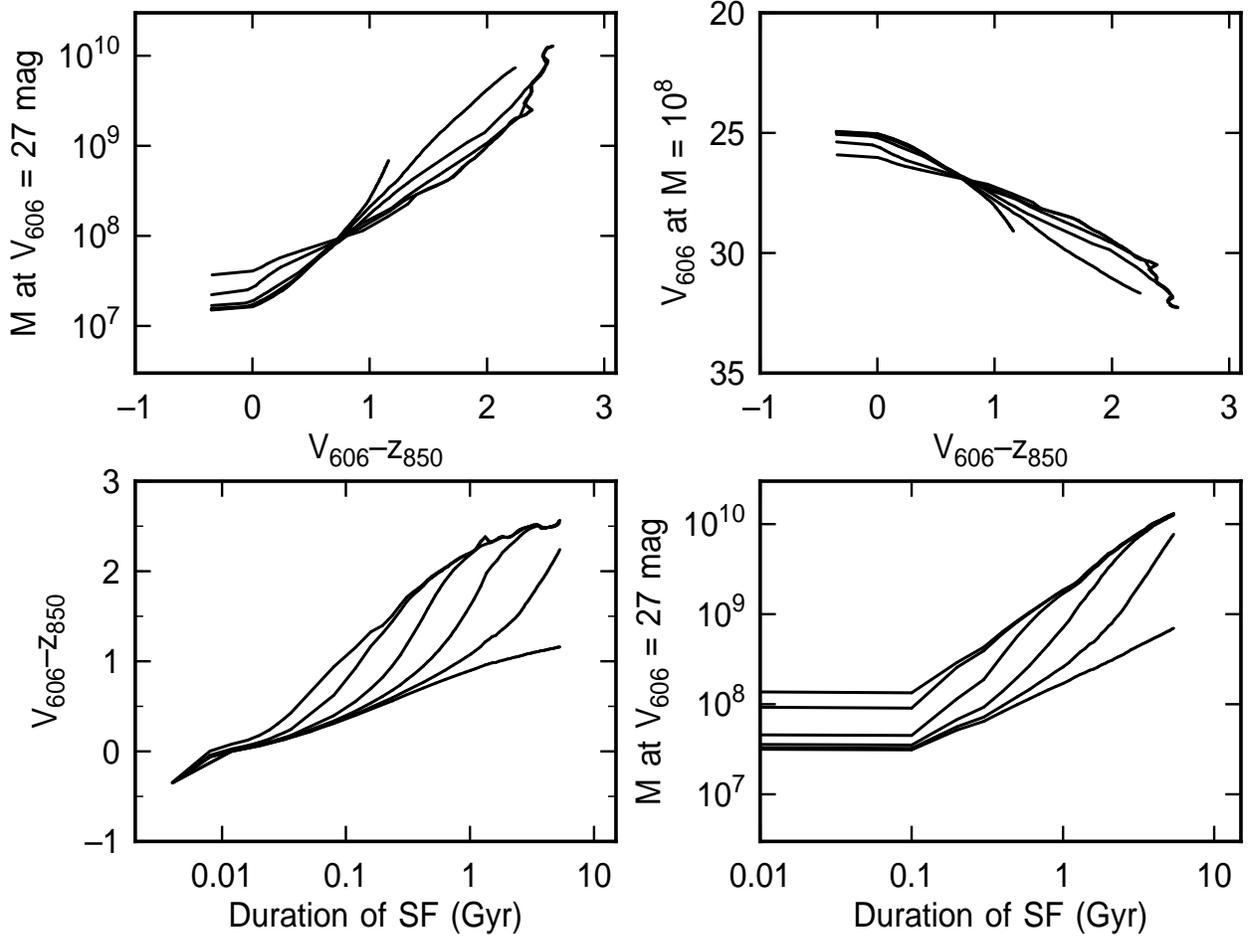}\caption{Models at $z=1$ for clump color (bottom
left) and clump mass at an apparent $V_{606}$ magnitude of 27 (lower
right) are shown in the bottom panels versus the duration of star
formation in 6 models with exponentially decaying star formation.
Five lines are for decay times of $10^7$, $3\times10^7$, $10^8$,
$3\times10^8$, and $10^9$ years, and the sixth line represents a
constant rate. Shorter decay times correspond to redder color (upper
lines) and higher masses (upper lines). In the top panels, the clump
mass at $V_{606}=27$ (top left) and the clump apparent magnitude at
$10^8$ M$_\odot$ masses (top right) are shown versus the clump
color. The correspondence between color and mass gives a degeneracy
to plots of mass versus color at a fixed apparent magnitude (top
left) and apparent magnitude versus color at a fixed mass (top
right). Thus the masses of clumps can be derived approximately from
their $V_{606}-z_{850}$ colors and $V_{606}$ magnitudes for each
redshift.}\label{fig:masscol0}\end{figure}

Figure \ref{fig:masscol0} shows sample model results for redshift
$z=1$. The different lines in each panel correspond to different
decay times for the star formation rate, in years: $10^7$,
$3\times10^7$, $10^8$, $3\times10^8$, and $10^9$, and the sixth line
represents a constant rate. Generally the shorter the decay time,
the redder the color and higher the mass for a given duration of
star formation. This correspondence between color and mass gives a
degeneracy to plots of mass versus color at a fixed apparent
magnitude (top left) and apparent magnitude versus color at a fixed
mass (top right). Thus the masses of clumps can be derived
approximately from their colors and magnitudes, without needing to
know their ages or star formation histories.

\begin{figure}\epsscale{.7}
\plotone{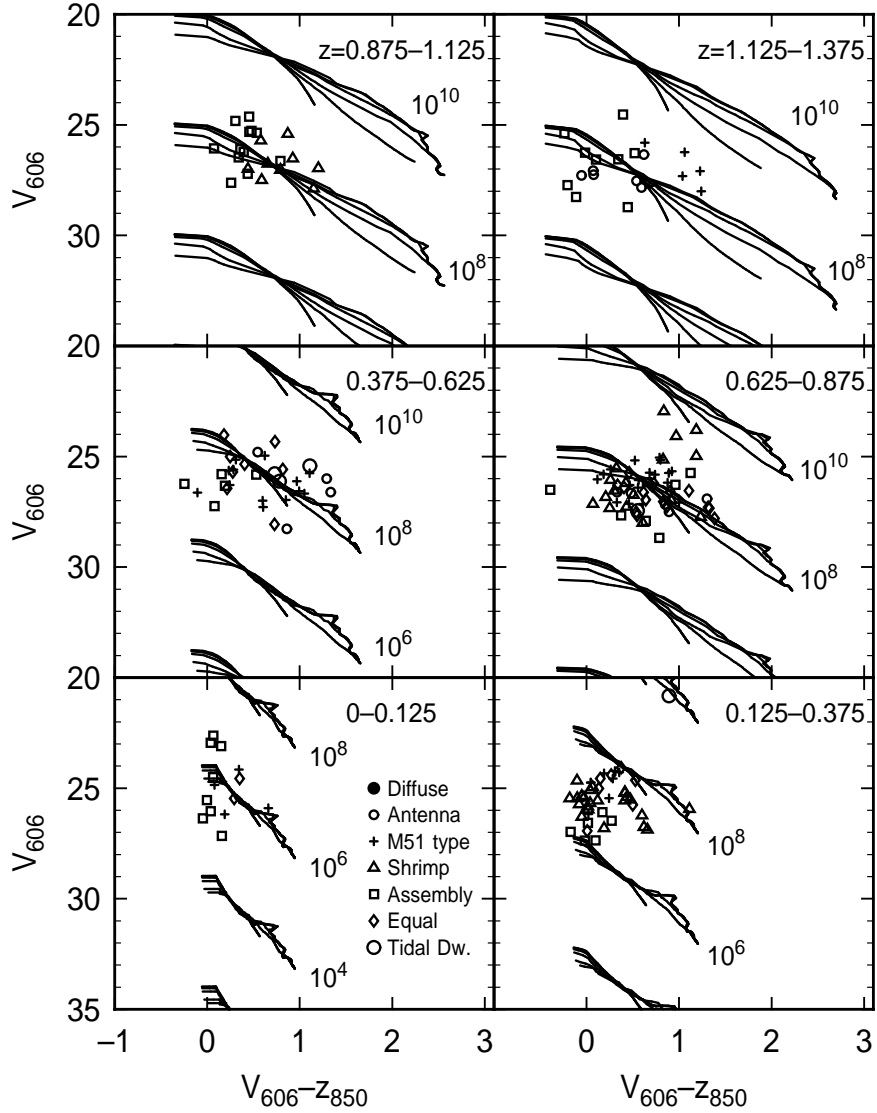}\caption{The masses of the clumps can be estimated
from this figure. Each curve in a cluster of curves is a different
model for color-magnitude evolution of a star-forming region, with
the age of the region changing along the curve and the exponential
decay rate of the star formation changing from curve to curve. The
different clusters of curves correspond to different total masses
for the star-forming regions (mass in M$_\odot$ is indicated to the
right of each curve). The symbols represent observations of apparent
magnitude and color. Bandshifting and absorption are considered by
plotting the observations and models in redshift bins. The mass
scales shift slightly with redshift. The mass of each star-forming
region can be determined by interpolation between the curves.
Typical masses are $10^6$ M$_\odot$ for low $z$ and $10^8$ M$_\odot$
for high $z$. The circle near the $10^{10}$ M$_\odot$ curves in the
$z=0.125-0.375$ interval corresponds to the diffuse clump in the
tidal debris of galaxy 1 in Fig. 1.
}\label{fig:masscol2}\end{figure}

Figure \ref{fig:masscol2} shows observations and models in the
color-magnitude plane for 6 redshift intervals spanning our
galaxies.  Each curve represents a wide range of star formation
durations that vary along the curve as in the top right panel of
Fig. \ref{fig:masscol0}; each curve in a set of curves is a
different decay time. The different sets of curves, shifted
vertically in the plots, correspond to different clump masses, as
indicated by the adjacent numbers, which are in M$_{\odot}$. Each
different point is a different clump; many galaxies have several
points. Only clumps with both $V_{606}$ and $z_{850}$ magnitudes
above the 2$\sigma$ noise limit are plotted in Figure
\ref{fig:masscol2}. The clump ($V_{606} - z_{850}$) colors range
from 0 to 1.5. The magnitudes tend to be about constant for each
redshift because of a selection effect (brighter magnitudes are rare
and fainter magnitudes are not observed).

\begin{figure}\epsscale{1.0}
\plotone{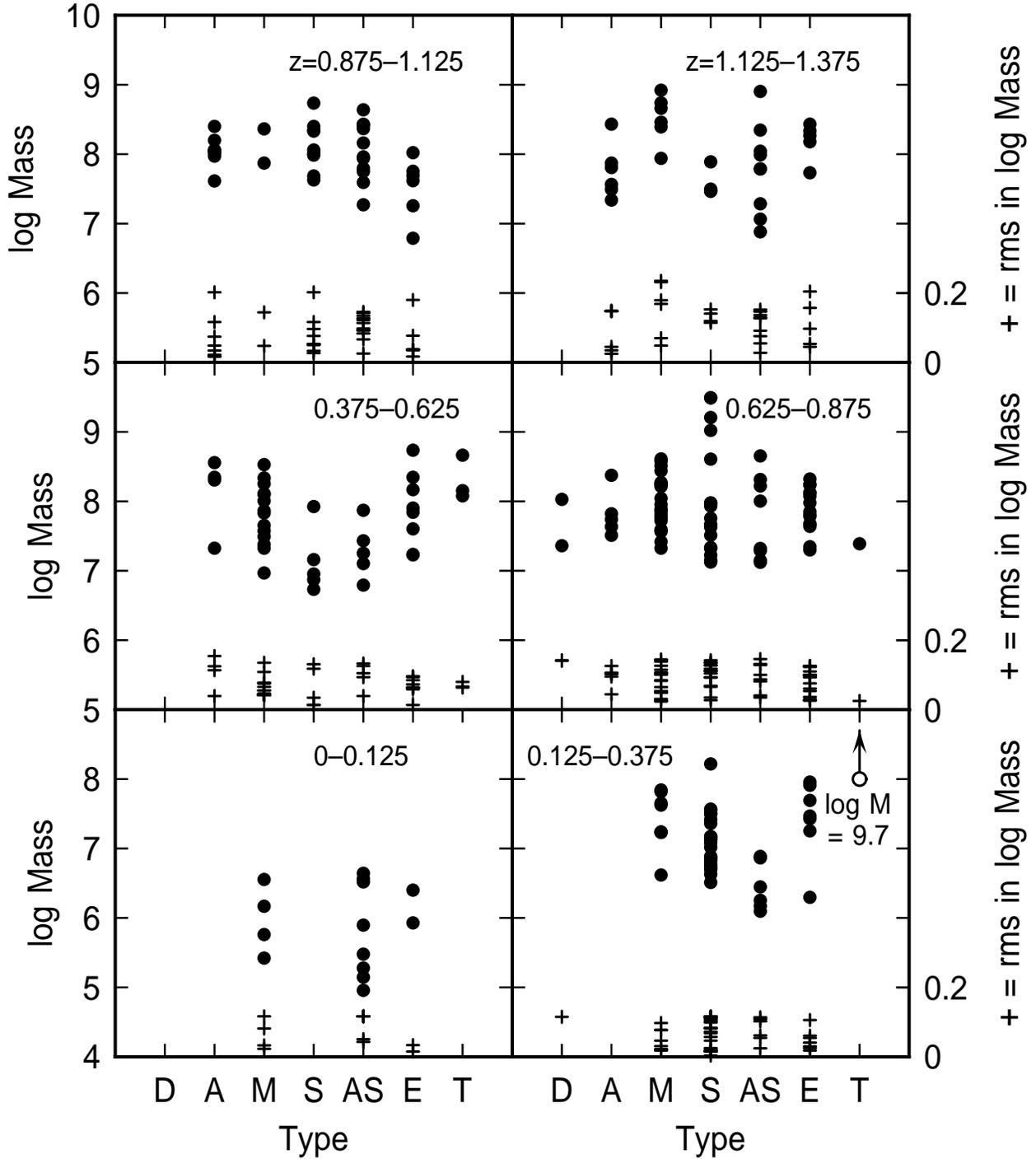}\caption{Clump masses (left axis) are plotted
versus galaxy type in order of Figs. 1-6: Diffuse, Antenna,
M51-type, Shrimp, Assembly, and Equal, with T representing the tidal
dwarfs. The method of Fig. 11 is used. The rms deviations among the
six star formation decay times are shown as plus-symbols using the
right-hand axes. }\label{fig:masscol3}\end{figure}

Figure \ref{fig:masscol2} indicates that the masses of the
observable clumps are between $10^6$ and $10^9$ M$_{\odot}$ for all
redshifts, with higher masses selected for the higher redshifts. The
masses for all of the clumps are plotted in Figure
\ref{fig:masscol3} versus the galaxy type (types 1 through 6 are in
order of Figs. 1 through 6 above). The masses are obtained from the
observed values of $V_{606}$ and $V_{606}-z_{850}$ using the method
indicated in Figure \ref{fig:masscol2}. The different mass
evaluations for the six decay times are averaged together in the log
to give the log of the mass plotted as a dot in Figure
\ref{fig:masscol3}. The rms values of log-mass among these six
evaluations are shown in Figure \ref{fig:masscol3} as plus-symbols,
using the right-hand axes. These rms deviations are less than 0.2,
so the uncertainties in star formation decay times and clump ages do
not lead to significant uncertainties in the clump mass. (Systematic
uncertainties involving extinctions, stellar evolution models,
photometric redshifts, and so on, would be larger.)

\begin{figure}\epsscale{1.0}
\plotone{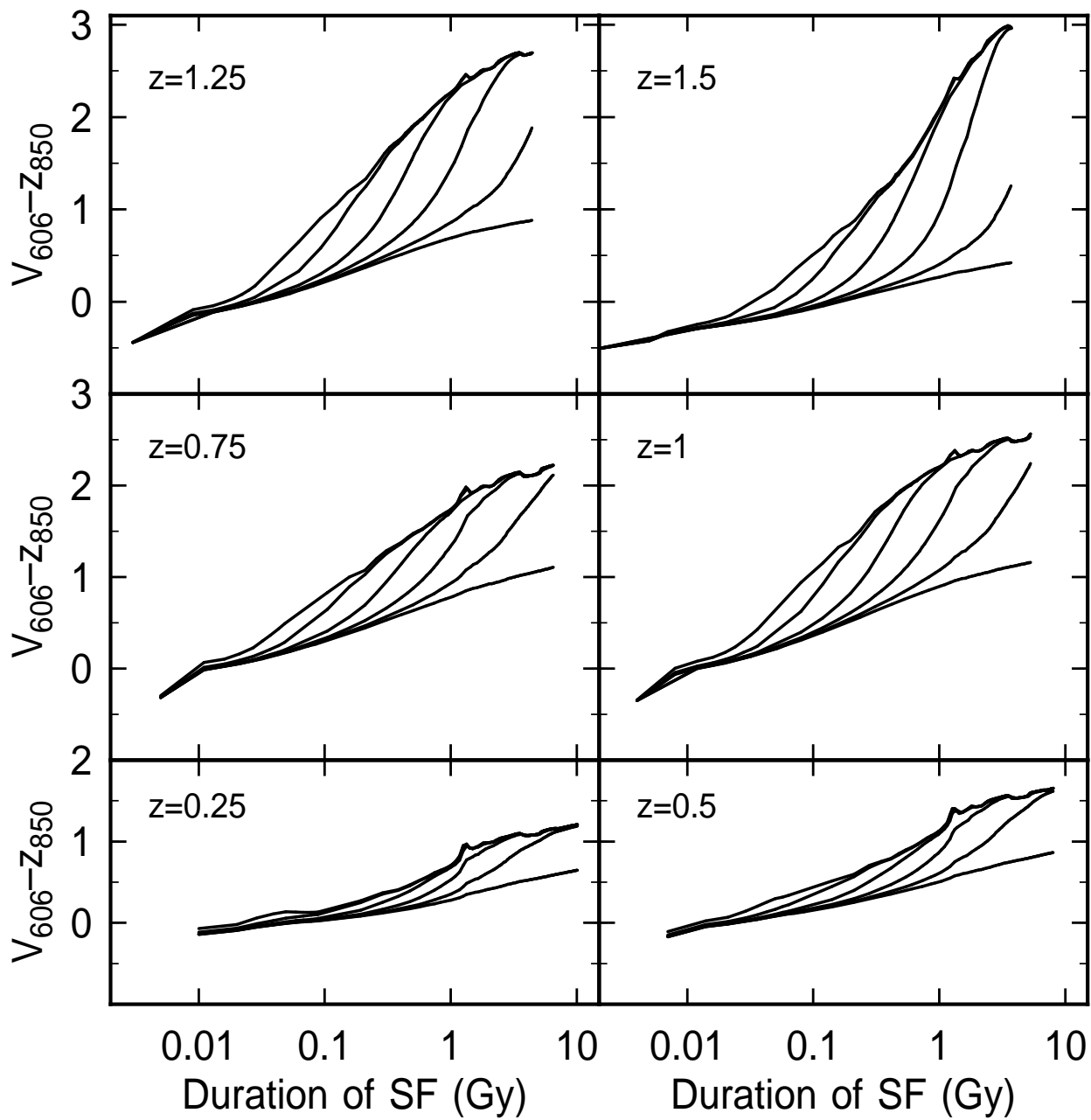}\caption{The apparent color of a star forming
region is shown versus the duration of star formation for an
exponentially decaying star formation law. The decay times are as in
Fig. 10, with short decay times the upper lines and continuous star
formation the lower lines. Using the observed clump colors, the
durations of star formation are found to range between $10^7$ and
$3\times10^8$ yrs for short decay
times.}\label{fig:masses2}\end{figure}

The clump ages cannot be determined independently from the star
formation decay times with only the few passbands available at high
angular resolution. Figure \ref{fig:masses2} shows model results
that help estimate the clump ages.  As in the other figures, each
line is a different exponential decay time for the star formation
rate. If we consider the two extreme decay times in this figure
(continuous star formation for the bottom lines in each panel and
$10^7$ years for the top lines), then we can estimate the age range
for each decay time from the observed color range. For
$V_{606}-z_{850}$ colors in the range from 0 to 0.5 at low $z$ (cf.
Fig \ref{fig:masscol2}), the clump ages range from $10^7$ to
$10^{10}$ yr with continuous star formation and from $10^7$ to
$3\times10^8$ yr with a decay time of $10^7$ yrs. For colors in the
range from 0 to 1.5 at higher redshifts, the age ranges are about
the same in each case. For intermediate decay times, the typical
clump ages are between $\sim10^7$ years for the bluest clumps and
$\sim10^9$ years for the reddest clumps. These are reasonable ages
for star formation regions, and consistent with model tail
lifetimes.

The star-forming complexes in the GEMS and GOODS interacting
galaxies are 10 to 1000 times more massive than the local analogs
seen in non-interacting late-type galaxies (Elmegreen \& Salzer
1999), but the low mass end in the present sample is similar to the
high mass end of the complexes measured in local interacting
galaxies. For example, the Tadpole galaxy, UGC 10214, contains
$10^6$ M$_{\odot}$ complexes along the tidal arm (Tran et al. 2003;
Jarrett et al. 2006). The interacting galaxy NGC 6872 has tidal
tails with $10^9$ M$_{\odot}$ HI condensations (Horellou \&
Koribalski 2007), but the star clusters have masses only up to
$10^6$ M$_{\odot}$ (Bastian et al. 2005). The most massive complexes
in the tidal tail of NGC 3628 in the Leo Triplet are also $\sim10^6$
M$_{\odot}$ (Chromey et al. 1998). The NGC 6872 clusters differ
qualitatively from those in our sample in being spread out along a
narrow arm; ours are big round clumps spaced somewhat evenly along
the arm. Small star clusters are also scattered along the tidal arms
the Tadpole and Mice systems; they typically contain less than
$10^6$ M$_{\odot}$ (de Grijs et al. 2003). The NGC 3628 clusters are
also faint with surface brightnesses less than 27 mag arcsec$^{-2}$;
they would not stand out at high redshift.

It is reasonable to consider whether the observed increase of
complex mass with increasing redshift is a selection effect. Our
clumps are several pixels in size, corresponding to a scale of
$\sim1$ kpc. Individual clusters are not resolved and we only sample
the most massive conglomerates. These kpc sizes are comparable to
the complex sizes in local galaxies, but the high redshift complexes
are much brighter and more massive. They would be observed easily in
local galaxies. The massive complexes in our sample are more similar
to those measured generally in UDF galaxies (Elmegreen \& Elmegreen
2005).

Clump separations were measured for clumps along the long arms in
the shrimp galaxies of Figure \ref{fig4shrimps}.  They average
$2.20\pm0.94$ kpc for 49 separations.  This is about the same
separation as that for the largest complexes in the spiral arms of
local spiral galaxies (Elmegreen \& Elmegreen 1983, 1987), and
comparable to the spacing between groups of dust-feathers studied by
La Vigne et al. (2006). Yet the clumps in shrimp galaxies and others
studied here are much more massive than the complexes in local
spiral arms, which are typically $<10^6$ M$_\odot$ in stars and
$\sim10^7$ M$_\odot$ in gas. This elevated mass can be explained by
a heightened turbulent speed for the gas, combined with an elevated
gas density.  Considering that the separation is about equal to the
two-dimensional Jeans length, $\lambda\sim2a^2/\left(G\Sigma\right)$
for velocity dispersion $a$ and mass column density $\Sigma$, and
that the mass is the Jeans mass, $\lambda^2\Sigma$, the mass scales
with the square of the velocity dispersion,
$M=M_0\left(a/a_0\right)^2$ for fixed length
$\lambda_0=2a_0^2/\left(G\Sigma_0\right)$ and
$M_0=\lambda_0^2\Sigma_0$. The mass column density also scales with
the square of the dispersion, $\Sigma=\Sigma_0\left(a/a_0\right)^2$
to keep $\lambda$ constant. Thus the interacting tidal arm clumps
are massive because the velocity dispersions and column densities
are high. Another way to derive this is to note that for regular
spiral arm instabilities, $2G\mu/a^2$ is about unity at the
instability threshold, where $\mu$ is the mass/length along the arm
(Elmegreen 1994). Thus cloud mass scales with $a^2$ for constant
cloud separation.  High velocity dispersions for neutral hydrogen,
$\sim50$ km s$^{-1}$, are also observed in local interacting
galaxies (Elmegreen et al. 1993; Irwin 1994; Elmegreen et al. 1995;
Kaufman et al. 1997; Kaufman et al. 1999; Kaufman et al. 2002).
Presumably the interaction agitates the interstellar medium to make
the large velocity dispersions. The orbital motions are forced to be
non-circular and then the gaseous orbits cross, converting orbital
energy into turbulent energy and shocks. Similar evidence for high
velocity dispersions was found in the masses and spacings of star
forming complexes in clump cluster galaxies (Elmegreen \& Elmegreen
2005) and in spectral line widths (Genzel et al. 2006; Weiner et al.
2006).

\subsection{Tail Properties}

\begin{figure}\epsscale{1.0} \plotone{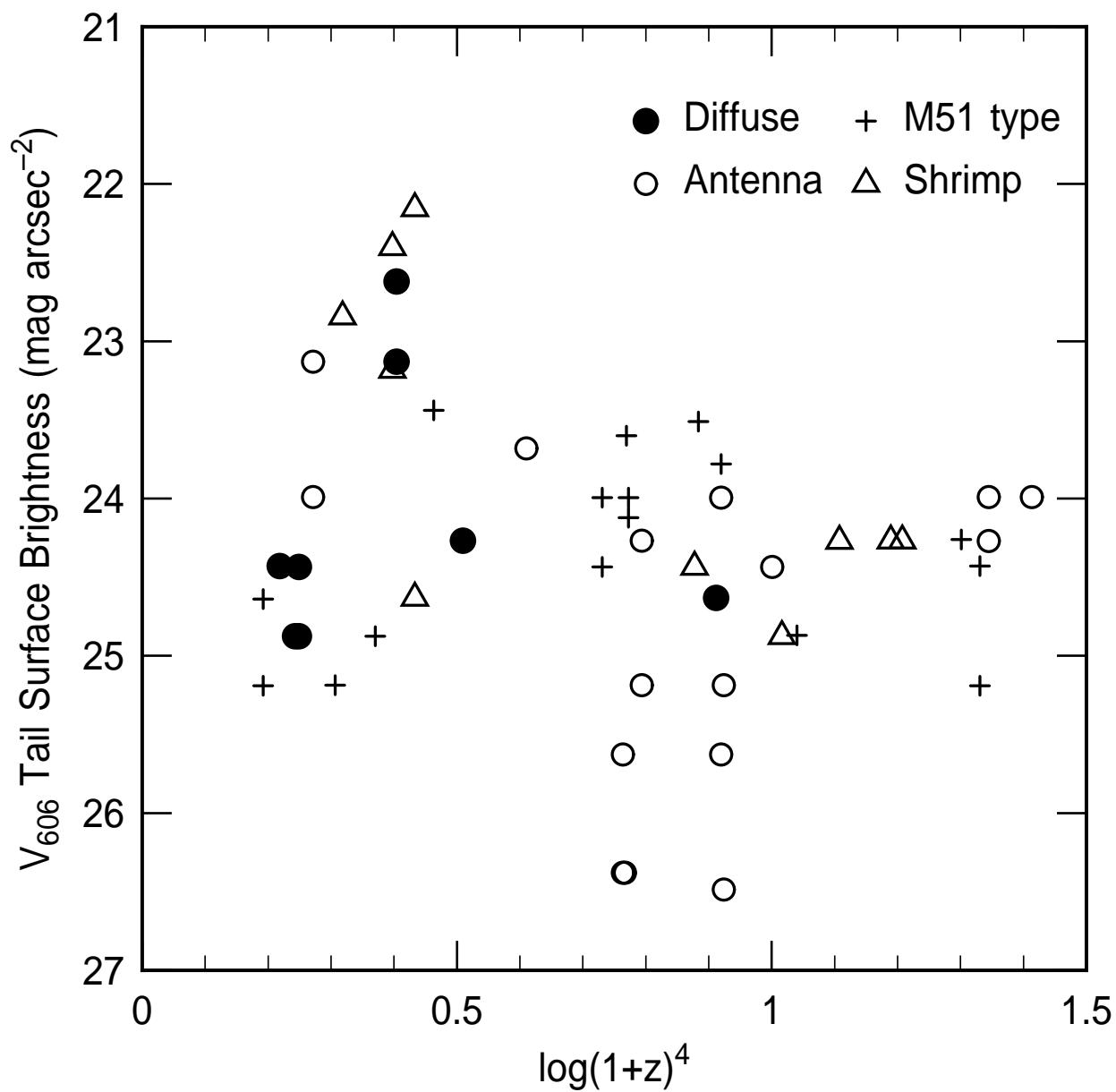}\caption{V-band
surface brightness of tidal tails for galaxies in Figures 1-4
plotted as a function of $(1+z)^4$ for redshift $z$. Some systems
have more than one tail. Cosmological dimming causes a decrease with
redshift equal to 2.5 magnitudes for each factor of 10 in
$\left(1+z\right)^{-4}$; this decrease is consistent with the
dimming seen here. The observable $2\sigma$ limit for these fields
is $\sim25$ mag arcsec$^{-2}$. Some antenna galaxies have patchy
tails with fainter average surface
brightnesses.}\label{fig:shrimf7}\end{figure}

Figure \ref{fig:shrimf7} shows the average tail surface brightness
as a function of $(1+z)^4$ for galaxies in Figures 1-4. Some systems
have more than one tail. Cosmological dimming causes a fixed surface
brightness to get fainter as $(1+z)^{-4}$, so there should be an
inverse correlation in this diagram. Clearly, the tails are brighter
for the more nearby galaxies, and they decrease out to $z\sim1$,
where they are fairly constant.  This constant limit is at the
2$\sigma$ detection limit of 25 mag arcsec$^{-2}$. Antennae galaxies
with average tail surface brightnesses fainter than this limit have
patchy tails with no apparent emission between the patches. Only the
brightest high redshift tails can be observed in this survey.

Simulations by Mihos (1995) suggested that tidal tails are
observable for a brief time in the early stages of a merger,
corresponding to $\sim150$ Myr at a redshift $z=1$ and 350 Myr at
$z=0.4$. The difference is the result of surface brightness dimming
as tails disperse. A nearby galaxy merger, Arp 299, has a 180 kpc
long tail encompassing 2 to 4\% of the total galaxy luminosity, with
an interaction age of 750 Myr, but its low surface brightness of
28.5 mag arcsec$^{-2}$ (Hibbard \& Yun 1999) would be below the
GOODS/GEMS detection limit.

\begin{figure}\epsscale{1.0} \plotone{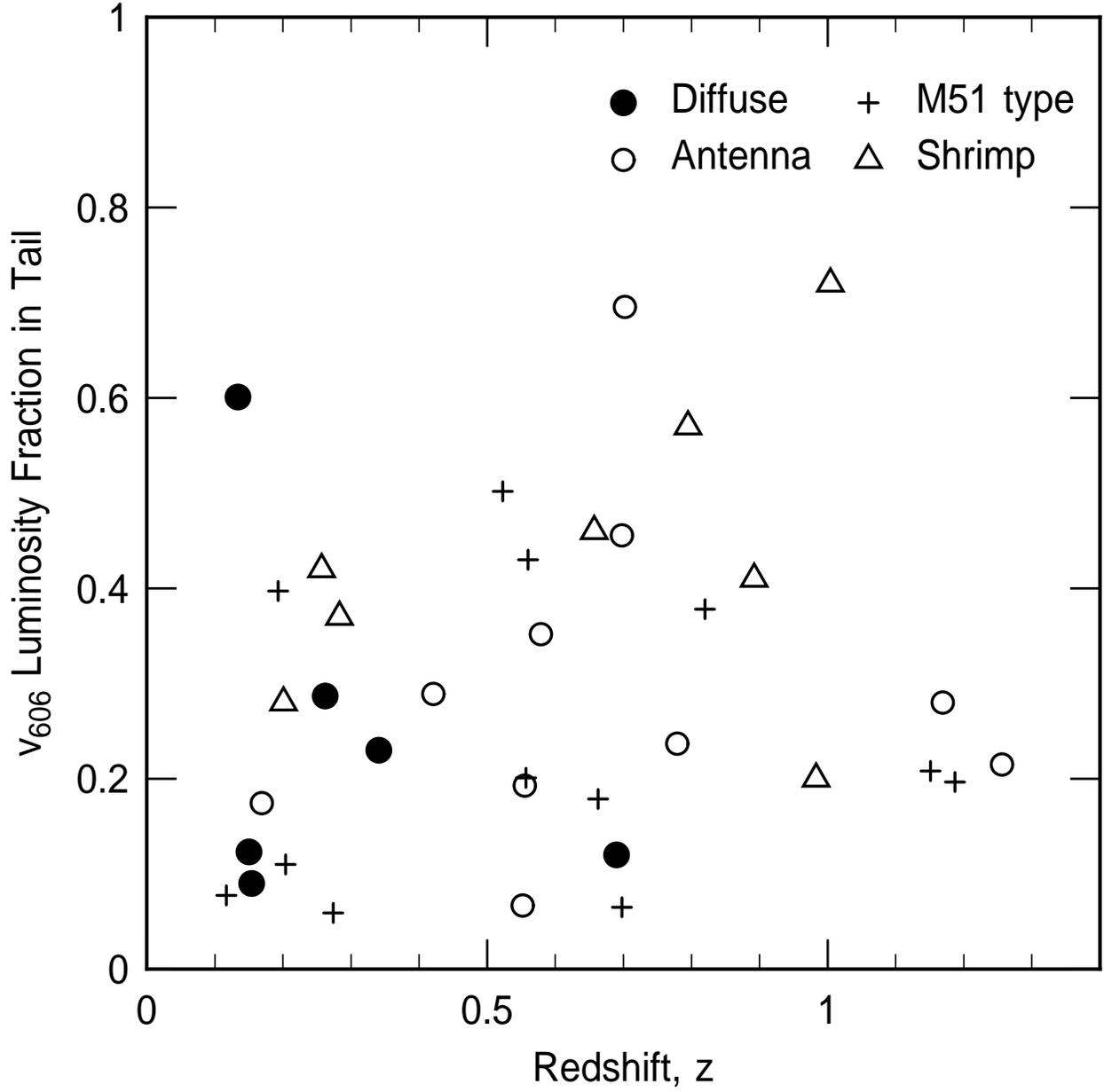}\caption{Fraction
of V-band luminosity in antennae tidal tails relative to their integrated
galaxy luminosity, as a function of redshift.}\label{fig:shrimf8}\end{figure}

The ratio of the luminosity of the combined tails and bridges to the
luminosity of the disk (the luminosity fraction) is shown in Figure
\ref{fig:shrimf8}. The luminosity fraction in the tidal debris
ranges from 10$\%$ to 80$\%$, averaging about 30$\%$ regardless of
redshift. This range is consistent with that of local galaxies in
the Arp atlas and Toomre sequence (e.g., Schombert, Wallin, \&
Struck-Marcell 1990; Hibbard \& van Gorkom 1996).

Interaction models with curled tails, as in our shrimp galaxies,
were made by Bournaud et al. (2003). Their models had dark matter
halos with masses $\sim10$ times the disk mass and extents less than
12 disk scale lengths. Some of our shrimp galaxies have one
prominent curved arm that is pulled out from the main disk but not
very far, resulting in a lopsided galaxy. Simulations indicate that
such lopsidedness may be the result of a recent minor merger
(Bournaud et al. 2005). In some of our cases, a nearby companion is
obvious.

\begin{figure}\epsscale{1.0} \plotone{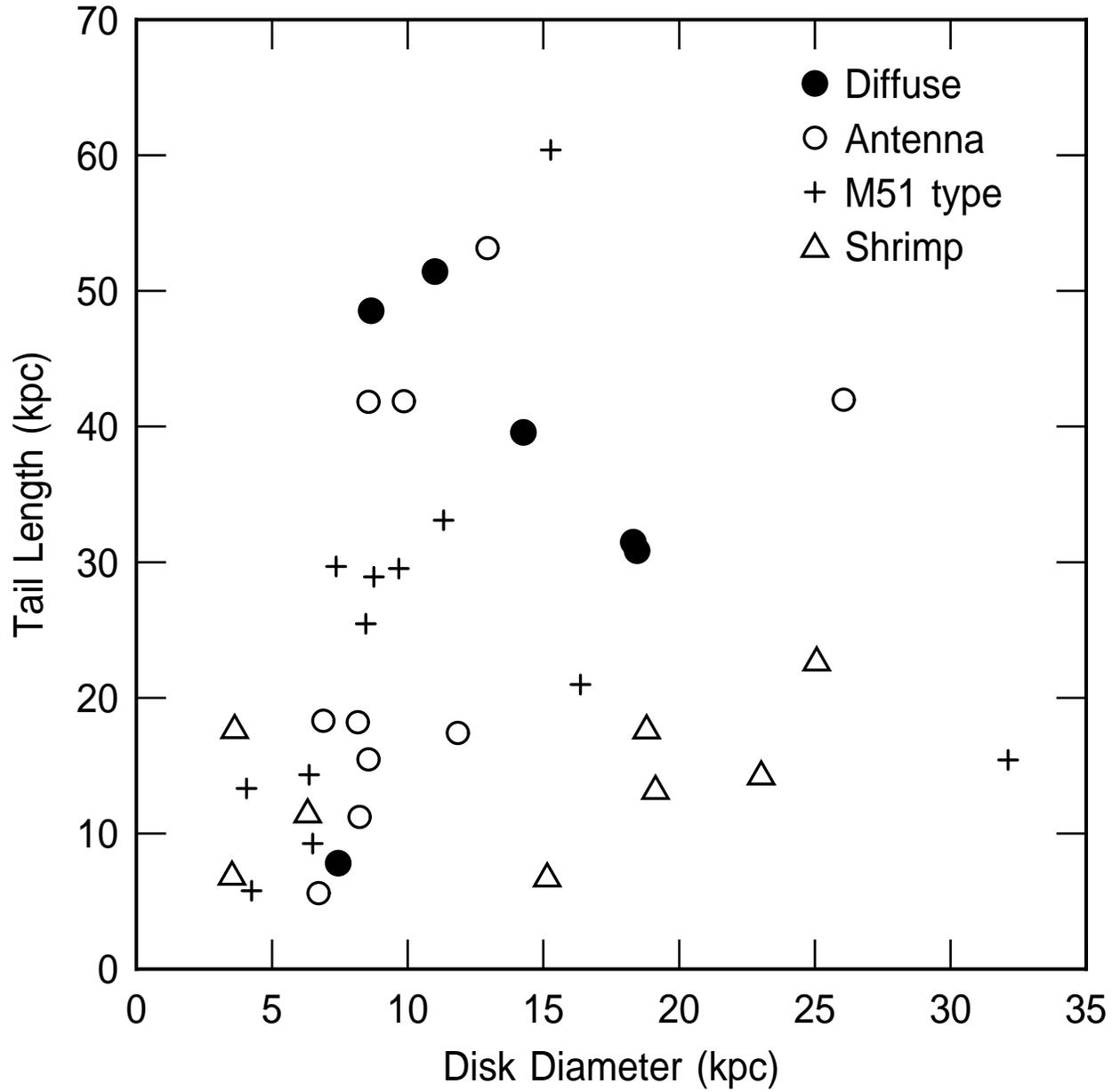}\caption{Tail
length versus disk diameter from Figs. 1-4, based on the V-band
images. Conversions to linear size assumed a standard $\Lambda$CDM
cosmology applied to the photometric
redshifts.}\label{fig:shrim10}\end{figure}

\begin{figure}\epsscale{1.0} \plotone{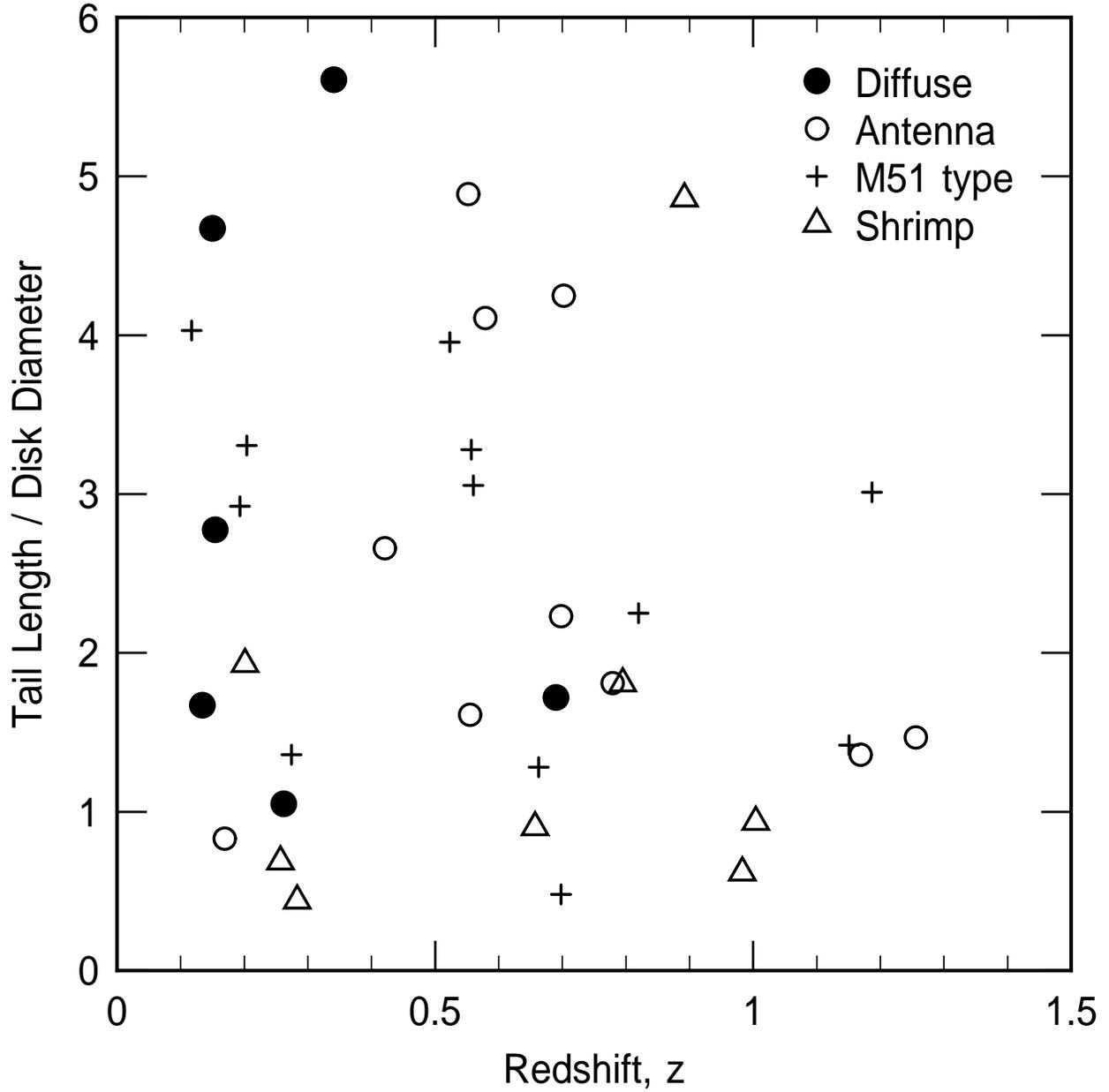}\caption{Tail
length/disk diameter as a function of redshift for shrimps and
antennae, measured from the V-band images. There is no obvious
trend. }\label{fig:shrim11}\end{figure}

The linear sizes of the tidal tails in our sample are shown in
Figure \ref{fig:shrim10}. They range from 2 to 60 kpc, and are
typically a few times the disk diameter, as shown in Figure
\ref{fig:shrim11}, which plots this ratio versus redshift. The
average tail to diameter ratio is $2.9\pm1.7$ for diffuse tails,
$2.5\pm1.3$ for antennae, $2.5\pm1.1$ for M51-types and $1.5\pm1.4$
for shrimps, so the shrimps are about 60\% as extended as the
antennae types.  There is no apparent dependence of these ratios on
redshift in Figure \ref{fig:shrim11}. Projection effects make these
apparent ratios smaller than the intrinsic ratios.

\begin{figure}\epsscale{1.0}\plotone{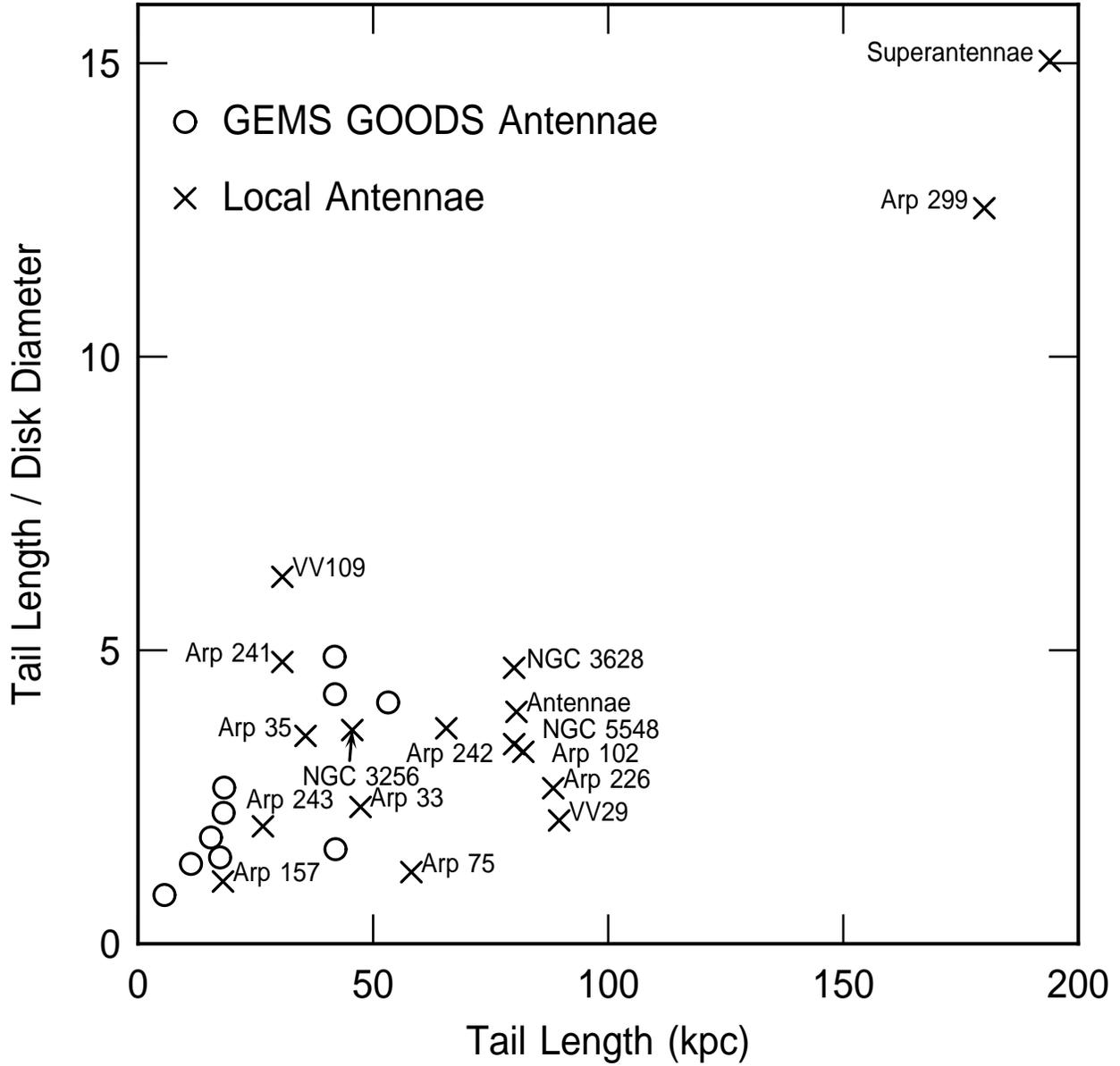}\caption{Tail
length/disk diameter versus the tail length for antenna galaxies in
our sample as well as for local antennae, whose names are indicated.
The GEMS and GOODS systems are significantly smaller than the local
antenna galaxies, even if the two extreme local cases, the
Superantennae and Arp 299, are excluded.
}\label{fig:shrimf16}\end{figure}

For comparison, the ratio of tail length to disk diameter versus the
tail length for local galaxies is shown in Figure \ref{fig:shrimf16}
based on measurements of antennae-type systems in the Arp atlas
(1966) and the Vorontsov-Velyaminov atlas (1959). Our galaxies are
also shown. The average tail length for the local galaxies in this
figure is $72\pm48$ kpc, while the average tail length for the GEMS
and GOODS antennae is 37\% as much, $27\pm16$ kpc. The diameters for
these two groups are $20\pm12$ kpc and $11\pm5$ kpc, and the ratios
of tail length to diameter are $4.5\pm3.7$ and $2.5\pm1.3$,
respectively.  Thus the local antennae mergers are larger in
diameter by a factor of 2 than the GEMS and GOODS antennae, and the
tails for the locals are larger by a factor of 2.7.  These results
for the diameters are consistent with other indicators that galaxies
are smaller at higher redshift, although usually this change does
not show up until $z>1$ (see observations and literature review in
Elmegreen et al. 2007).

\subsection{Tidal dwarf galaxy candidates}

Three antennae galaxies at the top of Figure \ref{fig2bants},
numbers 15, 16, and 18, have long straight tidal arms with large
star-forming regions at the ends. These clumps are possibly tidal
dwarf galaxies. The clump diameters and restframe B magnitudes are
listed in Table \ref{tab:dwarf}, along with the clump in diffuse
galaxy number 1 discussed in Sect. \ref{sect:sample}.  Listed are
their $V_{606}$ and $V_{606}-z_{850}$ magnitudes and associated
masses, calculated as in Sect. \ref{sect:clump}. The masses range
from $0.2\times10^8$ to $4.6\times10^8$ M$_\odot$ for the
star-forming dwarfs, but for the stellar condensation in the
diffuse-tail galaxy 1 (Fig. 1), the mass is $50\times10^8$
M$_{\odot}$. The star-forming dwarf masses are similar to or larger
than those found for the tidal object at the end of the
Superantennae (Mirabel et al. 1991) as well as the tidal object at
the end of the tidal arm in the IC 2163/NGC 2207 interaction
(Elmegreen et al. 2001) and at the end of the Antennae tail (Mirabel
et al. 1992). The HI dynamical masses for these local tidal dwarfs
are $\sim10^9$ M$_{\odot}$.

Simulations of interacting galaxies that form tidal dwarf galaxies
require long tails and a dark matter halo that extends a factor of
10 beyond the optical disk (Bournaud et al. 2003).  If one or both
galaxies contain an extended gas disk before the interaction, then
more massive, $10^9$ M$_{\odot}$ stellar objects can form at the tips of the
tidal arms from the accumulated pool of outer disk material
(Elmegreen et al. 1993; Bournaud et al. 2003). Observations of
nearby interactions show clumpy regions of tidal condensations with masses of $\sim10^8 -
10^9$ M$_{\odot}$ (Bournaud et al. 2004; Weilbacher et al. 2002,
2003; Knierman et al. 2003; Iglesias-Paramo \& Vilchez 2001), like
what is observed in our high redshift tidal dwarfs.

No well-resolved models have yet formed tidal dwarfs from stellar
debris. Wetzstein, Naab, \& Burkert (2007) considered this
possibility and found collapsing gas more likely. Yet the condensed
object in the tail of galaxy 1 could have formed there and it is
interesting to consider whether the Jeans mass in such an
environment is comparable to the observed mass.  If, for example,
the tidal arm surface density corresponds to a value typical for the
outer parts of disks, $\sim10$ M$_\odot$ pc$^{-2}$, and the stellar
velocity dispersion is comparable to that required in Sect.
\ref{sect:clump} for the gas to give the giant star forming regions,
$\sim40$ km s$^{-1}$, then the Jeans mass is $M\sim
a^4/\left(G^2\Sigma\right)\sim10^{10}$ M$_\odot$. This is not far
from the value we observe, $5\times10^9$ M$_\odot$, so the diffuse
clump could have formed by self-gravitational collapse of tidal tail
stars. The timescale for the collapse would be $a/\left(\pi
G\Sigma\right)\sim300$ Myr, which is not unreasonable considering
that the orbit time at this galactocentric radius is at least this
large.

\section{Dark Matter Halo Constraints}

Models of interacting galaxies have been used to place constraints
on dark halo potentials. Springel \& White (1999) and Dubinski,
Mihos, \& Hernquist (1999) found that tidal tail lengths can be long
compared to the disk if the ratio of escape speed to rotation speed
at 2 disk scale lengths is small, $v_e/V_r<2.5$, and the rotation
curve is falling in the outer disk. In a series of models, Dubinski
et al. showed that this condition may result from either
disk-dominated rotation curves where the halo is extended and has a
low concentration, or halo-dominated rotation curves where the halo
is compact and low mass. Dubinski et al. point out that the latter
possibility is inconsistent with observed flat or rising disk
rotation curves, but the first is compatible if the disk is massive
and dominant in the inner regions.  The first case also gives
prominent bridges.  In addition, Springel \& White (1999) found that
CDM halo models with embedded disks allow long tidal tails, but
Dubinski et al. noted that most of those which do are essentially
low surface brightness disks in massive halos, and not normal bright
galaxies. Galaxies without dark matter halos are not capable of
generating long tidal tails (Barnes 1988). In all cases, longer
tails develop in prograde interactions.

The smooth diffuse types and antenna types in Figures 1 and 2 have
relatively long tails, so the progenitors were presumably disks of
early and late types, respectively, with falling rotation curves in
their outer parts. These long-tail cases are relatively rare,
comprising only about 8\% and 9\%, respectively, of our original
(300 galaxy) interacting sample from GEMS and GOODS. The more
compact M51 types and shrimps represent 9\% and 12\% of the sample.
Short tail interactions could be younger, less favorably projected,
or have a more steeply rising rotation curve than long tail
interactions. The M51 types have clear companions, so the prominent
features are bridges. According to Dubinski et al. (1999), bridging
requires a prograde interaction with a maximum-disk galaxy, that is,
one with a low-mass, extended halo.

\section{Conclusions}

Mergers and interactions out to redshift $z=1.4$ have tails,
bridges, and plumes that are analogous to features in local
interacting galaxies. Some interactions have only smooth and red
features, indicative of gas-free progenitors, while others have
giant blue star-formation clumps. The tail luminosity fraction has a
wide range, comparable to that found locally. A striking difference
arises regarding the tail lengths, however. The tails in our antenna
sample, at an average redshift of 0.7, are only one-third as long as
the tails in local antenna mergers, and the disk diameters are about
half the local merger diameters. This difference is consistent with
the observations that high redshift galaxies are smaller than local
galaxies, although such a drop in size has not yet been seen for
galaxies at redshifts this low. The implication is that dark matter
halos have not built up to their full sizes for typical galaxies in
GEMS and GOODS.

Star formation is strongly triggered by the interactions observed
here, as it is locally.  The star-forming clumps tend to be much
more massive than their local analogs, however, with masses between
$\sim10^6$ M$_{\odot}$ and a few $\times10^8$ M$_\odot$, increasing
with redshift. This is not merely a selection effect, since the massive clumps seen at high redshift would show up at lower redshift, although of course smaller  clumps would not be resolved at high redshift. The clump spacings were measured along the tidal arms
of the most prominent one-arm type of interaction, the shrimp-type,
and found to be $2.20\pm0.94$, which is typical for the spacing
between beads on a string of star formation in local spiral arms. If
both types of arms form clumps by gravitational instabilities, then
the turbulent speed of the interstellar medium in the GEMS and GOODS
sample has to be larger than it is locally by a factor of $\sim5$ or
more; the gas mass column density has to be larger by this factor
squared.

Some interactions have tidal dwarf galaxies at the ends of their
tidal arms, similar to those found in the Superantennae galaxy and
other local mergers. One diffuse interaction with red stellar tidal
debris has a large stellar clump that may have formed by
gravitational collapse in a stellar tidal arm; the clump mass is
$5\times10^9$ M$_\odot$. Long-arm interactions are relatively rare,
comprising only $\sim17$\% of our total sample of $\sim300$
interacting systems (only a fraction of which were discussed here).
For those with long arms, numerical models suggest the dark matter
halos must be extended, so that the rotation curves are falling in
the outer disks. Most interactions are not like this, however, so
the rotation curves are probably still rising in their outer disks,
like most galaxies locally.

We gratefully acknowledge summer student support for B.M. and T.F.
through an REU grant for the Keck Northeast Astronomy Consortium
from the National Science Foundation (AST-0353997) and from the
Vassar URSI (Undergraduate Research Summer Institute) program.
D.M.E. thanks Vassar for publication support through a Research
Grant. We thank the referee for useful comments.
This research has made use of the NASA/IPAC Extragalactic
Database (NED) which is operated by the Jet Propulsion Laboratory,
California Institute of Technology, under contract with the National
Aeronautics and Space Administration.

\clearpage

\begin{deluxetable}{lcccc}
\tabletypesize{\scriptsize} \tablewidth{0pt} \tablecaption{Interacting
Galaxies in GEMS and GOODS \label{table2}} \tablehead{
\colhead{Type, Figure} &\colhead{Number} & \colhead{COMBO 17} & \colhead{z} & \colhead{R mag.}\\
} \startdata
Diffuse (Fig. 1) & 1   &   6423    &   0.15    &   16.572  \\
&2   &   12639  &   0.154   &   16.678  \\
&3   &   11538  &   0.134   &   17.713  \\
&4   &   53129  &   0.171   &   16.968  \\
&5   &   57881  &   0.118   &   17.552  \\
&6   &   28509  &   0.093   &   18.79   \\
&7   &   17207  &   0.69    &   19.742  \\
&8   &   30824  &   0.341   &   19.755  \\
&9   &   25874  &   0.262   &   19.757  \\
Diffuse (other) & 10  &   22588   &   0.684   &   21.263  \\
&11  &   21990  &   0.429   &   21.243  \\
&12  &   46898  &   0.617   &   20.794  \\
&13  &   49709  &   0.302   &   20.23   \\
&14  &   15233  &   0.304   &   18.857  \\
Antennae (Fig. 2) &15  &   61546   &   0.552   &   20.41  \\
&16  &   45115  &   0.579   &   21.275  \\
&17  &   20280  &   0.555   &   21.653  \\
&18  &   41907  &   0.702   &   22.66   \\
&19  &   35611  &   1.256   &   22.655  \\
&20  &   10548  &   0.698   &   22.43   \\
&21  &   33650  &   0.169   &   18.86   \\
&22  &   42890  &   0.421   &   20.68   \\
&23  &   49860  &   1.169   &   23.632  \\
&24  &   34926  &   0.779   &   -19.69  \\
Antennae (other) & 25  &   14829   &   0.219   &   21.429  \\
&26  &   18588  &   0.814   &   22.748  \\
&27  &   46738  &   1.204   &   20.65   \\
&28  &   7551   &   1.162   &   25.926  \\
&29  &   20034  &   1.326   &   21.932  \\
&30  &   33267  &   0.067   &   23.112  \\
&31  &   38651  &   0.988   &   23.89   \\
&32  &   55495  &   1.00    &   24.261  \\
M51-type (Fig. 3)& 33  &   5640    &   0.204   &   19.477  \\
&34  &   9415   &   0.523   &   21.16   \\
&35  &   40901  &   0.193   &   19.751  \\
&36  &   17522  &   0.82    &   23.103  \\
&37  &   6209   &   1.187   &   22.723  \\
&38  &   23667  &   1.151   &   23.514  \\
&39  &   37293  &   0.274   &   20.533  \\
&40  &   39805  &   0.557   &   20.089  \\
&41  &   53243  &   0.698   &   21.683  \\
&42  &   15599  &   0.56    &   21.381  \\
&43  &   25783  &   0.663   &   20.732  \\
&44  &   39228  &   0.117   &   18.031  \\
M51-type (other)  & 45  &   1984    &   0.762   &   22.855  \\
&46  &   2760   &   1.281   &   23.202  \\
&47  &   15040  &   0.667   &   22.392  \\
&48  &   18502  &   0.228   &   21.942  \\
&49  &   14959  &   0.306   &   19.581  \\
&50  &   16023  &   0.668   &   21.887  \\
&51  &   30226  &   0.509   &   22.689  \\
&52  &   40744  &   0.292   &   21.119  \\
&53  &   45102  &   0.857   &   22.514  \\
&54  &   60582  &   0.946   &   22.54   \\
Shrimp (Fig. 4)  &  55  &   40198   &   0.201   &  20.55   \\
&56  &   14373  &   0.795   &   23.183  \\
&57  &   12222  &   1.004   &   22.417  \\
&58  &   28344  &   0.257   &   19.509  \\
&59  &   56284  &   0.657   &   21.667  \\
&60  &   2385   &   0.283   &   21.334  \\
&61  &   54335  &   0.892   &   22.824  \\
&62  &   28841  &   0.673   &   20.971  \\
&63  &   6955   &   0.983   &   22.24   \\
Shrimp (other)  &   64  &   34244   &   0.999   &  22.504  \\
&65  &   48298  &   0.429   &   21.663  \\
&66  &   37809  &   0.357   &   20.667  \\
&67  &   25316  &   0.985   &   23.717  \\
&68  &   49595  &   0.663   &   21.939  \\
&69  &   59467  &   0.487   &   21.568  \\
&70  &   9062   &   0.854   &   23.82   \\
&71  &   30076  &   0.832   &   22.672  \\
&72  &   2760   &   1.281   &   23.202  \\
&73  &   54335  &   0.892   &   22.824  \\
Assembly (Fig. 5) & 74  &   28751   &   0.093   &   23.506  \\
&75  &   4728   &   0.702   &   22.799  \\
&76  &   23187  &   1.183   &   23.565  \\
&77  &   45309  &   1.061   &   22.916  \\
&78  &   41835  &   0.098   &   19.134  \\
&79  &   61945  &   1.309   &   21.813  \\
&80  &   62605  &   1.011   &   23.143  \\
&81  &   44956  &   0.506   &   22.494  \\
&82  &   4546   &   0.809   &   22.163  \\
&83  &   23000  &   0.132   &   22.951  \\
&84  &   63112  &   0.499   &   22.273  \\
&85  &   43975  &   1.059   &   22.878  \\
Equal (Fig. 6) &  86  &   40813   &   0.182   &   19.983  \\
&87  &   8496   &   0.354   &   22.415  \\
&88  &   13836  &   0.661   &   21.054  \\
&89  &   11164  &   0.464   &   19.351  \\
&90  &   39877  &   0.493   &   22.142  \\
&91  &   40598  &   0.263   &   20.128  \\
&92  &   51021  &   0.743   &   20.96   \\
&93  &   35317  &   0.671   &   20.755  \\
&94  &   56256  &   0.502   &   20.309  \\
&95  &   47568  &   0.649   &   20.206  \\
&96  &   40766  &   0.46    &   19.997  \\
&97  &   24927  &   0.524   &   19.647  \\
&98  &   15233  &   0.304   &   18.857  \\
&99  &   18663  &   1.048   &   24.011  \\
&100 &   43242   &   0.657   &   21.177  \\
\enddata

\end{deluxetable}

\clearpage
%\newpage
\begin{table}
\begin{center}
\caption{Tidal Dwarf Galaxy Candidates\label{tab:dwarf}}
\end{center}
\begin{tabular}{ccccccccc}
\tableline\tableline Galaxy &   z  &Galaxy&   Diam. &Dwarf
&$V_{606}$&$V_{606}-z_{850}$&Clump Mass\\
(COMBO17 \#)&&$M_{B,rest}$ (mag)&(kpc)&$M_{B,rest}$ (mag)&mag&mag&x10$^8$ M$_{\odot}$\\
\tableline
1 (6423) &   0.15 &-20.64&   13.9    &   -17.55 &20.83&0.90        & 50    \\
15 (61546) &   0.552&-20.77   &   5.5    &   -16.67 &26.12&0.78       &1.2     \\
16 (45115)  &   0.579&-20.17   &   4.7    &   -17.72 &25.42&1.1        &4.6     \\
18 (41907)  &   0.702&-19.21   &   1.9    &   -16.03 &27.45&0.55       &0.24    \\
17 (20280)&0.555&-19.56&6.2&-17.06&25.77&0.73&1.4\\
\end{tabular}
\end{table}
% 1=8.18, 15=56.02, 16=34.02, 18=45.13, 17=20.15

\clearpage
%\newpage
%fig1

\clearpage
%fig2

\clearpage
%fig3

\clearpage
%fig4

\clearpage
%fig5

\clearpage
%fig6

\clearpage
%fig7

\clearpage
%fig8 ! redone still needs one wreck galaxy 25.15 xx

\clearpage
%fig9

\clearpage
%fig10
 \clearpage

\clearpage
%fig11
 \clearpage

\clearpage
%fig12
 \clearpage

\clearpage
%fig13

\clearpage
%fig14

\clearpage
%fig15

\clearpage
%fig16
 \clearpage

\clearpage
%fig17

\clearpage
%fig18

\end{document}